\documentclass{article}

\usepackage{microtype}
\usepackage{graphicx}
\usepackage{subcaption}
\usepackage{booktabs} 

\usepackage{hyperref}

\usepackage[accepted]{icml2026}

\usepackage{amsmath}
\usepackage{amssymb}
\usepackage{mathtools}
\usepackage{amsthm}

\usepackage{booktabs}
\usepackage{siunitx}
\usepackage{multirow}
\usepackage{enumitem}
\usepackage{pifont}
\usepackage{bbm}

\usepackage{algpseudocode}
\usepackage{tabularx}
\usepackage{makecell}
\algrenewcommand\algorithmicindent{2em}

\newcommand{\COMMENT}[1]{\textit{$\triangleright$ #1}}

\usepackage[capitalize,noabbrev]{cleveref}

\theoremstyle{plain}
\newtheorem{theorem}{Theorem}[section]

\newtheorem{lemma}[theorem]{Lemma}
\newtheorem{corollary}[theorem]{Corollary}
\theoremstyle{definition}
\newtheorem{definition}[theorem]{Definition}

\theoremstyle{remark}

\usepackage[textsize=tiny]{todonotes}

\icmltitlerunning{ZipMoE: Efficient On-Device MoE Serving via Lossless Compression and Cache-Affinity Scheduling}

\begin{document}

\twocolumn[
\icmltitle{
ZipMoE: Efficient On-Device MoE Serving via Lossless Compression and Cache-Affinity Scheduling
}




\icmlsetsymbol{equal}{*}

\begin{icmlauthorlist}
\icmlauthor{Yuchen Yang}{equal,njuese}
\icmlauthor{Yaru Zhao}{equal,njuese}
\icmlauthor{Pu Yang}{njuese}
\icmlauthor{Shaowei Wang}{njuese}
\icmlauthor{Zhi-Hua Zhou}{njukeysoft,njuai}
\end{icmlauthorlist}

\icmlaffiliation{njuese}{School of Electronic Science and Engineering, Nanjing University, China.}
\icmlaffiliation{njukeysoft}{National Key Laboratory for Novel Software Technology, Nanjing University, China.}
\icmlaffiliation{njuai}{School of Artificial Intelligence, Nanjing University, China}

\icmlcorrespondingauthor{Shaowei Wang}{wangsw@nju.edu.cn}


\icmlkeywords{On-device LLM, Machine Learning Systems, Scheduling, Mixture-of-Experts, Edge Computing, ICML}

\vskip 0.3in
]



\printAffiliationsAndNotice{\icmlEqualContribution}

\begin{abstract}

While Mixture-of-Experts (MoE) architectures substantially bolster the expressive power of large-language models, their prohibitive memory footprint severely impedes the practical deployment on resource-constrained edge devices, especially when model behavior must be preserved without relying on lossy quantization.
In this paper, we present \textsc{ZipMoE}, an efficient and \emph{semantically lossless} on-device MoE serving system.
\textsc{ZipMoE} exploits the synergy between the hardware properties of edge devices and the statistical redundancy inherent to MoE parameters via a caching-scheduling co-design with provable performance guarantee.
Fundamentally, our design shifts the paradigm of on-device MoE inference from an I/O-bound bottleneck to a \emph{compute-centric} workflow that enables efficient parallelization.
We implement a prototype of \textsc{ZipMoE} and conduct extensive experiments on representative edge computing platforms using popular open-source MoE models and real-world workloads.
Our evaluation reveals that \textsc{ZipMoE} achieves up to $72.77\%$ inference latency reduction and up to $6.76\times$ higher throughput than the state-of-the-art systems.
Our code is available at: \url{https://github.com/npnothard/ZipMoE-ICML26}.

\end{abstract}


\section{Introduction}
\label{sec:introduction}

Recent years have witnessed increasing interest in deploying large language models (LLMs) on edge and mobile devices.
On-device inference obviates the need to transmit sensitive data to remote clouds, improves service availability, and avoids the service interruption caused by the unreliable wide-area networks\cite{EdgeOPT, OSSR}, thereby enabling a broad class of mobile AI applications, including embodied intelligence \cite{humanoid}, personal assistants \cite{assistant}, and chatbots \cite{chatbot}.
Among existing LLM architectures, Mixture-of-Experts (MoE), a classic ensemble learning method \cite{Zhou1, Zhou2, AMLE}, has been found well effective in training modern LLMs.
By decomposing the model into specialized experts and activating only a small subset per token, MoE substantially increases model capacity while keeping computational overhead sub-linear to the model size.
However, the sparse activation nature of MoE models fundamentally shifts the system bottleneck from computation to memory, which presents a critical barrier to their practical on-device deployment under tight memory constraints.


To improve memory efficiency for on-device LLM inference, prior works have primarily focused on model quantization \cite{GPTQ, QwT} and pruning \cite{LLMPrune}, which reduce model size by representing model weights or structures in lower-precision or more compact forms.
Building on the insight that quantization sensitivity is highly non-uniform, varying across tensors \cite{FloE, MxMoE}, experts \cite{EdgeMoE}, and even tokens within the same expert \cite{D2MoE}, state-of-the-art solutions adopt fine-grained, adaptive bit-width allocation combined with system-aware optimizations to balance the accuracy–efficiency trade-off \cite{FlexGen}.
While effective in reducing memory footprint, such \emph{lossy} compression fundamentally shifts model behavior beyond what existing evaluation metrics (e.g., perplexity and zero-shot accuracy) can capture, particularly regarding security \cite{LLMVulnerable, QuantHurt}.
It is proved feasible to construct adversarial LLMs with quantized parameters that induce vulnerable code generation, over-refusal, and content injection attacks \cite{InstructionTuning}, whereas their full-precision counterparts behave benignly \cite{ExploitLLMQ}.
These findings highlight that quantization-based inference systems can introduce vulnerabilities to malicious attacks, particularly in unmonitored on-device deployments.

An alternative line of work exploits the inherent activation sparsity of MoE models by caching a subset of experts in GPU memory while offloading inactive parameters to lower-tier storage.
These offloaded tensors are fetched either proactively via data-driven predictors \cite{SiDAMoE, ProMoE} or reactively upon gate activation.
To mitigate I/O bottlenecks caused by frequently transferring experts over PCIe, existing serving systems employ pipelined execution to overlap GPU inference with I/O events \cite{MoELightning, Klotski}.
Despite improving throughput on consumer- and server-grade hardware, these approaches exhibit fundamental mismatches with real-world on-device deployment.
First, on-device applications (e.g., interactive chatbots and assistants) prioritize low-latency interactive inference where the batch size is typically one \cite{SwapMoE}, severely limiting the efficacy of pipelining and batch-level optimizations.
Second, existing systems operate on the premise that CPU main memory serves as a distinct external storage for offloaded tensors \cite{MixtralOffload, FineMoE, MoEInfinity}. 
This assumption rarely holds true and is rendered impractical on edge platforms (e.g., Jetson platforms, mobile phones, and Raspberry Pi) that are predominantly built on mobile system-on-chips (SoCs), where CPUs and GPUs share the same physical memory pool and bus resources.
Consequently, existing solutions fail to exploit, or even account for the unique performance characteristics of shared-memory architectures. 
Note that this oversight also severely undermines emerging CPU-GPU hybrid inference systems \cite{KTransformers, Fiddler}.
These fundamental mismatches between existing MoE serving systems and practical on-device deployments motivate us to reconsider a core question:

\emph{How can MoE models be efficiently served on mobile and edge platforms without altering their algorithmic behavior?}

In this paper, we tackle this challenge by presenting \textsc{ZipMoE}, the \emph{first} system to accelerate MoE inference on mobile and edge computing platforms through lossless compression.
Recognizing the significant statistical redundancy within specific bit-fields of MoE parameters, we first introduce a compression-decompression pipeline to obtain parameters that (i) minimize PCIe I/O through \emph{lossless} compression, and (ii) support parallel decompression on the CPU. 
This pipeline enables bit-level operations on tensor entries and ensures high efficiency by leveraging a zero-copy paradigm and a memory-coalesced GPU kernel for tensor recovery, which incurs negligible runtime overhead by fully exploiting the unified memory architecture (UMA) of mobile SoCs.
Second, we explore hierarchical, differentiated caching to manage tensors across distinct compression states for fine-grained memory control. 
We devise a cache-pool planning algorithm based on probabilistic modeling of expert skewness and dynamic programming to determine the optimal memory budget for each compression state.
This hybrid caching strategy enables parallel, asynchronous on-demand decompression, orchestrated by our cache-affinity scheduling algorithm, which we prove to perform within a constant factor of the global optimum.
Ultimately, our design rethinks expert loading in MoE inference by \emph{computing the required expert tensors} rather than \emph{blocking on memory transfers}, effectively unleashing the potential of the powerful yet often underutilized multi-core CPUs on modern SoCs.


\section{Background and Motivation}
\label{sec:background_and_motivation}

\subsection{System-Hardware Mismatch in Existing Offloading Solutions}

\begin{figure}
\centering
\includegraphics[scale=0.45]{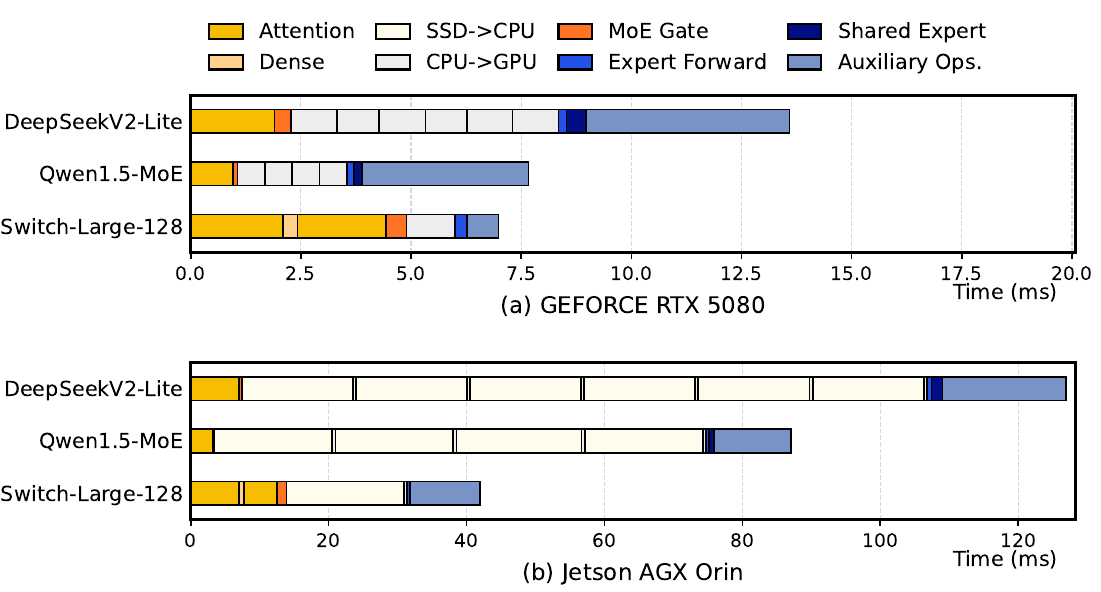}
\caption{Latency break-down of decoding layers in representative MoE models on (a) Server environment, where experts are offloaded to CPU with $512$ GB RAM; and (b) Edge environment, where experts are offloaded to NVMe SSD (Aigo DP35) with $2$ GB/s read speed.}
\label{fig:microbench1}
\end{figure}

Unlike server-grade systems, mainstream mobile and edge computing platforms (e.g., NVIDIA Jetson, Raspberry Pi, Apple Silicon) are predominantly UMA-based hardware, where  heterogeneous processors (e.g., CPUs and GPUs) share a single, coherent physical memory pool.
To examine the implications of this architectural shift, we benchmark offloading-based MoE inference on both a server platform (Fig.\hyperref[fig:microbench1]{1 (a)}) and an edge device (Fig.\hyperref[fig:microbench1]{1 (b)}).
The results reveal that existing offloading paradigms are ill-suited for edge hardware, primarily due to the following degradations:\\
\noindent
\textbf{Exacerbated I/O Stalls.}
Due to the memory capacity limitation in UMA-based mobile platforms, on-device MoE is characterized by intensive read operations from NVMe SSDs rather than CPU memory, whose read bandwidth  (typically $1-5$ GB/s) is significantly lower than that of server-grade CPU memory ($16-32$ GB/s).
Consequently, inference speed is severely bottlenecked by storage I/O.
As shown in Figure \ref{fig:microbench1}, I/O stalls surge from $38.5\%$ to $80.1\%$ for decoder-only models and from $15.8\%$ to $41.7\%$ for encoder-decoder models when moving from server to edge settings.\\
\noindent
\textbf{Underutilized Compute Power.}
Although UMA theoretically accelerates in-memory operations, as evidenced by the fact that edge environment offers a $2.12\times$ speedup in host-to-device transfers in our benchmark, this benefit is negated by the excessive expert I/O latency on the critical path of MoE inference.
Under typical on-device workloads with batch size one, prolonged I/O stalls leave both CPU and GPU resources largely idle, resulting in severe underutilization of the available compute capacity.

These findings demonstrate a fundamental mismatch between existing MoE serving systems and the hardware characteristics of mobile SoCs.

\subsection{Information-Inefficiency in MoE Parameters}


\begin{figure}
\centering
\includegraphics[scale=0.27]{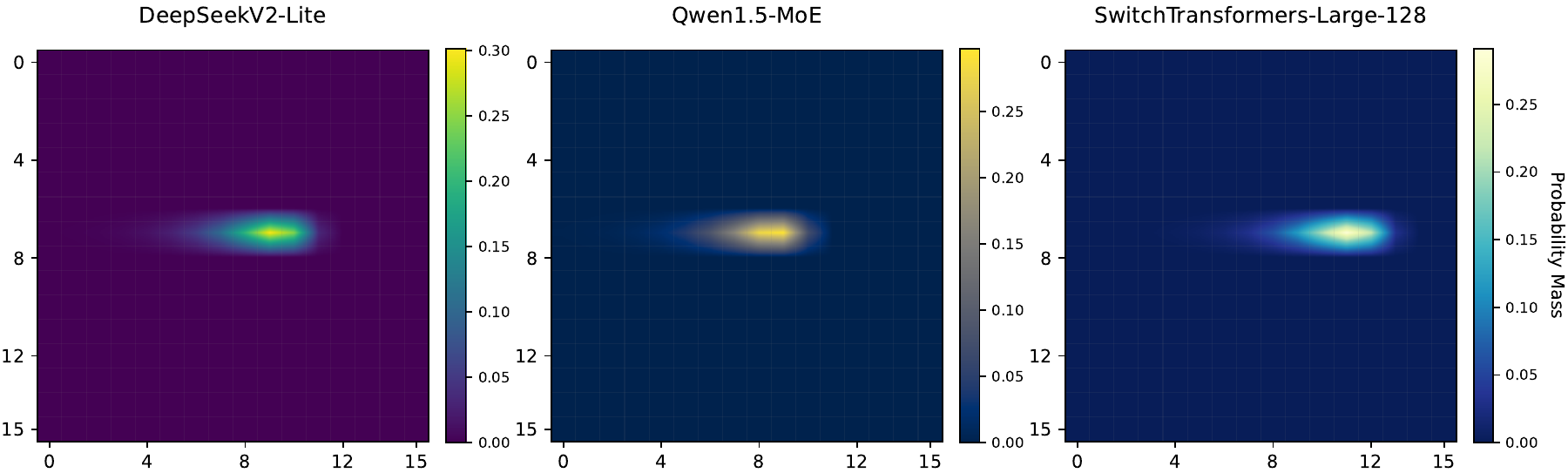}
\caption{The probability mass heatmaps of integer representations of the exponent bits extracted from different MoE parameters.
The Shannon entropy values for the three models are $2.651$ bits, $2.563$, and $2.554$ bits, respectively.}
\label{fig:entropy}
\end{figure}

\begin{figure}
\centering
\includegraphics[scale=0.365]{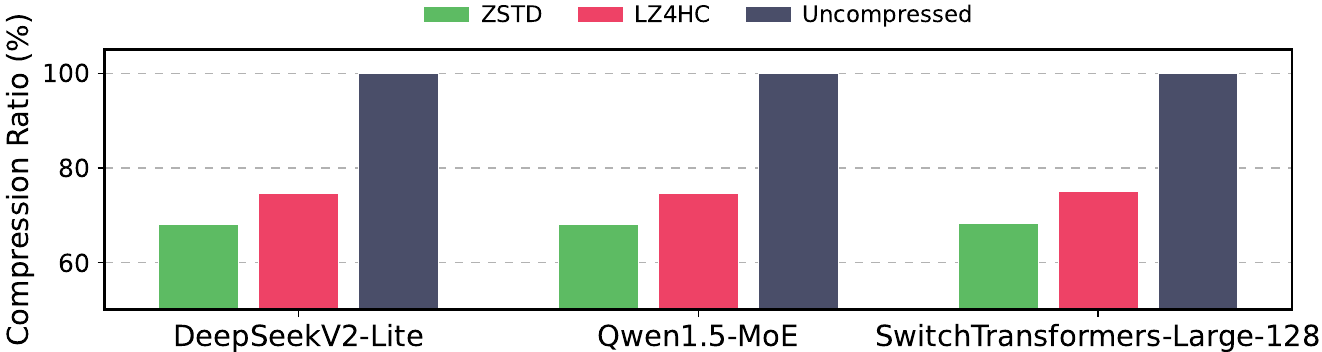}
\caption{The compression ratios of different MoE parameters using different lossless compressors.}
\label{fig:compression_ratio}
\end{figure}

The current \emph{de facto} Brain Floating Point (BF16) representation used by modern LLMs is information-theoretically inefficient, which burdens the I/O efficiency.\\
\noindent
\textbf{BF16 Parameters.}
In the BF16 format, each real number is represented by a \emph{sign} bit, 8 \emph{exponent} bits, and 7 \emph{mantissa} bits, and can be computed as $(-1)^{\text{sign}}\times 2^{\text{exponent}-127}\times(1.\text{mantissa})$.
The well-balanced nature between numerical range (up to $\pm3.39\times10^{38}$) and precision (down to $1.18\times10^{-38}$) has made BF16 widely adopted in both LLM training \cite{BF16Train} and inference \cite{GiveMeBF16}. \\
\noindent
\textbf{Low-Entropy Exponent Bits.}
Prior studies have observed pronounced entropy heterogeneity across different bit-fields in BF16 parameters \cite{ZipNN, HuffLLM, NeuZip, DF11}, where sign and mantissa bits exhibit near-random distributions, while exponent bits are highly redundant.
To quantify this effect in MoE models, we analyze the expert network parameters and plot the probability mass heatmaps of the exponent-bit symbols.
As shown in Fig.\ref{fig:entropy}, the distributions of exponent symbols are consistently skewed across all evaluated models, with the support set containing only $14.06\%$, $11.33\%$, and $14.84\%$ of symbols, respectively.
Such strong skewness directly indicates substantial statistical redundancy and high compressibility of BF16 representations from an information-theoretic perspective.
We further apply two general-purpose lossless compressors, LZ4HC \cite{LZ4} and ZSTD \cite{ZSTD}, to MoE parameters.
The normalized compressed sizes are shown in Fig. \ref{fig:compression_ratio}.
LZ4HC reduces model size to $74\%$, while ZSTD achieves $68\%$, approaching the Shannon entropy lower bound (calculated using numerical results in Fig.\ref{fig:entropy} as $66.02\%$, $65.96\%$, and $66.52\%$ for the three models, respectively).
These results indicate that current off-the-shelf compressors can already yield substantial size reduction.\\
\noindent
\textbf{Lossless Compression for LLM Inference.}
Despite its clear potential for reducing storage and I/O overhead, lossless compression remains largely under-explored in LLM inference, particularly for MoE models.
Existing approaches either target specialized hardware such as FPGAs \cite{HuffLLM}, or rely on nvCOMP \cite{nvCOMP}, which offers little support for AArch64 CPUs prevalent in mobile and edge platforms.
While DFloat11 \cite{DF11} enables on-the-fly GPU decompression for general LLMs, its computational overhead becomes prohibitive when it cannot be amortized across large batches, which is rarely encountered in on-device inference.
Most notably, existing lossless compression schemes are agnostic to the conditional activation patterns of MoE models and fail to address the scalability challenges posed by inevitable parameter offloading under tunable memory budgets.
This limitation is particularly critical on mobile and edge platforms, where the operating system manages multi-tenant workloads and system memory is shared across concurrent applications.

\subsection{From I/O-Bound to CPU-Parallel Expert Access}

\begin{figure}
\centering
\includegraphics[scale=0.68]{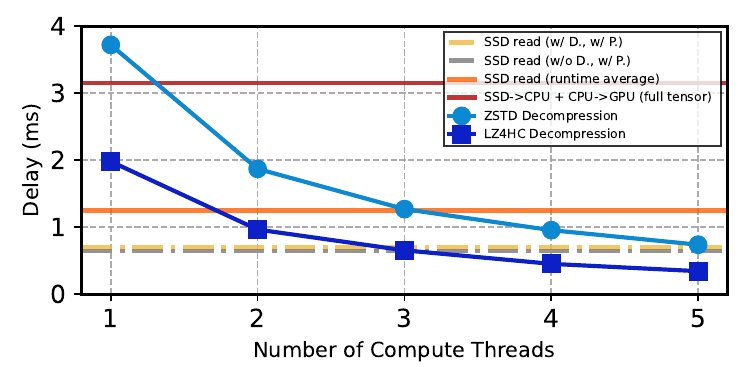}
\caption{Comparison between decompression delay and I/O delay in various cases, where D. denotes the existence of decompression in other threads, P. denotes page cache is enabled. All read operations are tested to read the same amount of bytes with decompressed tensors (the size of exponent bits), while full tensor represents the end-to-end delay of transferring the complete tensor to GPU.
All experiments are carried out on Jetson AGX Orin 64G with Samsung 970 EVO SSD.}
\label{fig:microbench3}
\vspace{-0.41cm}
\end{figure}

\begin{figure*}[t]
    \centering
    \includegraphics[width=0.98\textwidth]{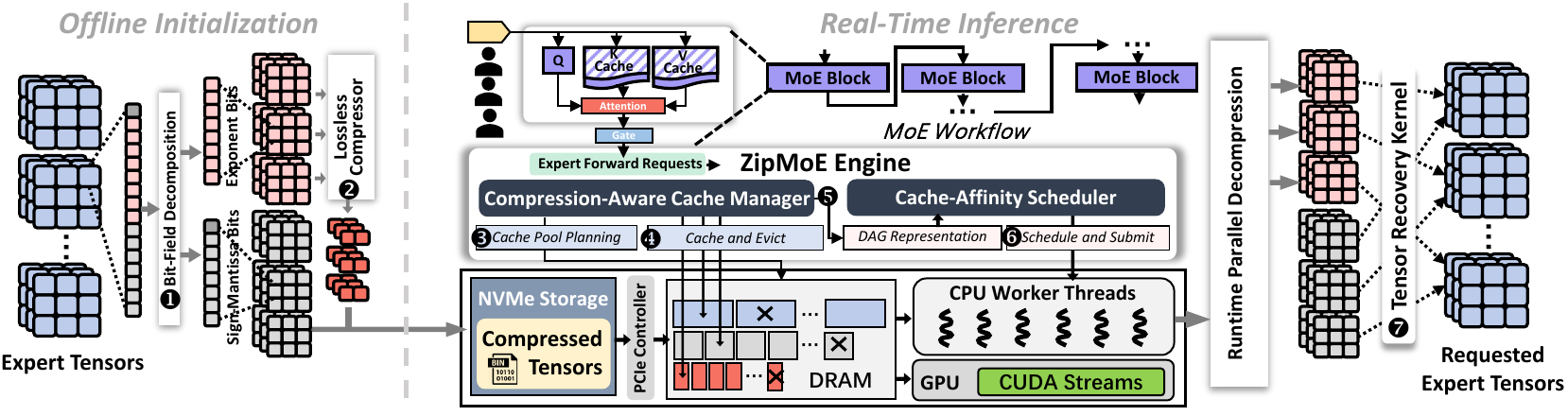}
    \caption{\textsc{ZipMoE} System Overview}
    \label{fig:system}
    \vspace{-0.2cm}
\end{figure*}

The above findings demonstrate that compressing offloaded parameters is a promising approach to reduce I/O traffic.
Motivated by the observation that modern mobile platforms are equipped with multi-core CPUs, we benchmark the decompression throughput of LZ4HC and ZSTD against raw tensor I/O.
Our results in Fig. \ref{fig:microbench3} reveal the following insights, which motivate the design principles of \textsc{ZipMoE}.\\
\noindent
\textbf{(\textrm{I}) Decompression is Not on the Critical Path.}
When reading the same number of bytes as the decompressed tensor payload, multi-threaded decompression incurs lower latency than SSD I/O once $\geq3$ worker threads are employed.
This indicates that decompression can be efficiently parallelized across CPU cores.
Moreover, since the exponent bits occupy the same storage footprint as the sign and mantissa bits combined, decompression latency can be effectively hidden by overlapping it with the loading these incompressible bits without introducing additional latency overhead.\\
\noindent
\textbf{(\textrm{II}) Resource Interference is Negligible.}
Since SSD reads and decompression-induced DRAM writes compete for the same memory controller bandwidth, we evaluated the impact of parallel background decompression on I/O performance.
Our measurements show a modest $7.47\%$ reduction in peak SSD read throughput.
We note that this overhead is negligible in practice.
As shown in Fig. \ref{fig:microbench3} (orange line), the average runtime SSD throughput is significantly lower than the peak bandwidth due to system-level factors (e.g., low page-cache hit rates).
Consequently, the contention at the memory controller does not become a performance bottleneck, implying that decompression can safely proceed in parallel with I/O without causing severe interference.\\
\noindent
\textbf{(\textrm{III}) Opportunities in Compression-Aware Caching.}
Recall that effective decompression overlap is achieved with as few as $3$ worker threads and yields even more speedups over full tensor read.
Meanwhile, mainstream mobile devices typically feature $6$–$12$ CPU cores.
This imbalance indicates that substantial CPU compute capacity remains underutilized during MoE inference.
Our key insight is that, since sign and mantissa bits are largely incompressible and constitute $50\%$ of the full tensor size, selectively caching them enables $2\times$ tensor coverage under the same memory budget.
Upon a partial cache hit, redundant CPU threads can recover the compressed exponent bits in parallel, with the decompression latency fully hidden by the I/O of other cache-missed experts, thereby achieves the same zero latency as a full-tensor cache hits.
Note that the same principle also applies when caching compressed exponent bits.
These observations motivate a hierarchical caching design that is explicitly compression-aware, enabling I/O reduction beyond the theoretical lower bound of entropy.

\section{ZipMoE Design}
\label{sec:design}

\subsection{System Overview}

Fig. \ref{fig:system} illustrates the architecture of \textsc{ZipMoE}, which consists of two stages: \emph{offline initialization} and \emph{real-time inference}.\\
\noindent
\textbf{Offline Initialization.}
Executed once prior to deployment, this stage converts model parameters into lossless compressed representations.
\textsc{ZipMoE} first performs bit-field decomposition (Fig. \ref{fig:system} \ding{202}) to separate tensor elements into low-entropy exponent bits and high-entropy sign–mantissa bits.
The extracted exponent bits are partitioned into $K$ shards and then compressed (denoted as \texttt{E-chunks}), while the sign–mantissa bits are packed into byte-aligned representations (denoted as \texttt{SM-chunks}).
All components, together with the required metadata, are serialized into a binary format and offloaded to NVMe SSD. (Fig. \ref{fig:system} \ding{203})\\
\noindent
\textbf{Real-time Inference.}
To start up, the \emph{compression-aware cache manager} determines the cache capacity allocated to tensors at different compression states, i.e. the compressed \texttt{E-chunks}, \texttt{SM-chunks}, and fully reconstructed tensors, under a given memory budget and execution configuration (Fig. \ref{fig:system} \ding{204}).
At runtime, once the gate network reveals the selected experts, corresponding expert requests are issued to the \emph{cache-affinity scheduler}, which dispatches them to appropriate cache pools based on their access patterns (Fig. \ref{fig:system} \ding{205}).
For each request, \textsc{ZipMoE} constructs a cache-aware directed acyclic graphs (DAGs) that captures the required I/O, decompression, and reconstruction operations (Fig. \ref{fig:system} \ding{206}).
These operations are scheduled to execute asynchronously across $L$ parallel CPU worker threads, enabling overlapped \texttt{SM-chunk} loading and \texttt{E-chunk} decompression (Fig. \ref{fig:system} \ding{207}).
Finally, the decompressed tensors are reassembled into BF16 format using a memory-coalesced GPU kernel for model inference (Fig. \ref{fig:system} \ding{208}).

\subsection{Compression States}
\label{sec:compression_states}

\begin{figure}
\centering
\includegraphics[scale=0.39]{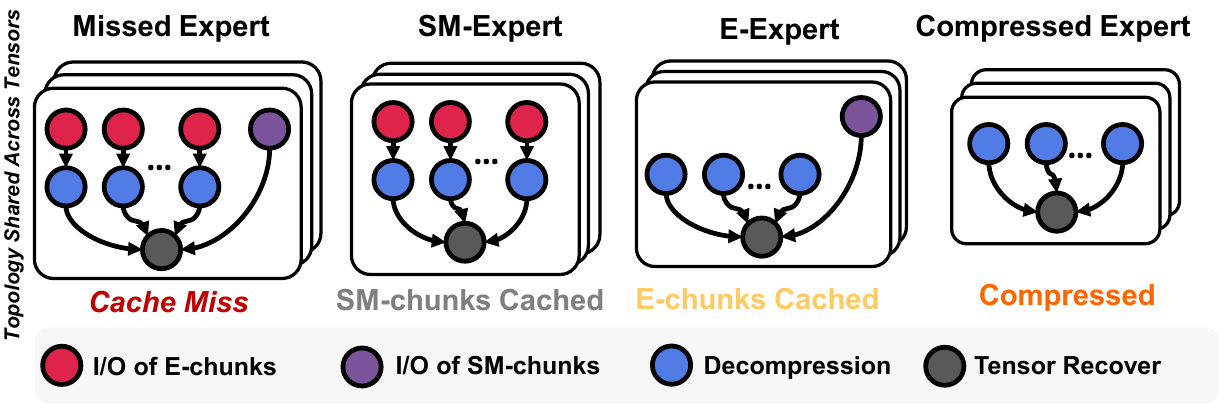}
\caption{DAG structures of expert requests with different compression states.}
\label{fig:DAG}
\end{figure}

\textsc{ZipMoE} achieves fine-grained memory management at runtime by allowing the \texttt{E-chunks}, \texttt{SM-chunks}, and full tensors of each expert to be cached separately.
To this end, we introduce the \emph{compression state} abstraction to track the residency of these components in memory. 
Specifically, an expert is defined as an \texttt{E-expert} if its \texttt{E-chunks} are cached, and an \texttt{SM-expert} if its \texttt{SM-chunks} are cached. 
An expert is termed as \texttt{compressed-expert} when both components reside in memory.\\
\noindent
\textbf{DAG Representation.}
To enable adaptive resource scheduling, each expert reconstruction task is modeled as a DAG, whose nodes represent the fine-grained operations such as decompression, SSD read, and tensor recovery, and the directed edges denote the precedence constraints.
\textsc{ZipMoE} dynamically constructs the DAG topology for each task based on the expert’s runtime compression state.
Note that the task granularity is defined at the tensor level.
For an expert comprising $N$ tensors, all tensors share the same topology and generate $N$ independent reconstruction tasks.
We show all possible DAG structures in Fig. \ref{fig:DAG}.

\subsection{Cache-Affinity Scheduler}
\label{sec:scheduler}

Once the selected experts are revealed, a set $\mathcal{Q}$ of DAG tasks are issued to the scheduler.
Let $\rho$ denote the compression ratio of the lossless compressor used by \textsc{ZipMoE}, which is defined as the size of compressed exponent bits relative to their original representation.
For each task $j\in\mathcal{Q}$, we denote the I/O latency of a single \texttt{SM-chunk} as $u$.
Accordingly, reading one compressed \texttt{E-chunk} incurs an I/O cost of $\frac{\rho}{K} u$, and its corresponding decompression cost is denoted as $c$.
By $n(j)$ we denote the expert ID corresponding to task $j$.
For each expert $n$, let $p_n$ represent the GPU execution time required to process all tokens routed to that expert.
We create a dedicated I/O thread for all SSD reads, a pool of $L$ CPU worker threads for decompression, and a CUDA stream for tensor recovery.
Our objective is to minimize the makespan of each sparse layer, i.e., the completion time of the last expert execution.

We design an efficient scheduling algorithm to orchestrate the fine-grained operations in the execution of every sparse MoE layer, aiming to maximize parallelism across heterogeneous resources while minimizing idle time caused by I/O stalls. \\
\noindent
\textbf{Task Partition.}
Based on their critical execution paths, we partition all expert tasks into two disjoint classes.
\emph{Type-I tasks} correspond to experts whose reconstruction requires loading \texttt{SM-chunks}, and therefore incur expensive blocking on the I/O thread.
\emph{Type-II tasks} include the remaining experts whose \texttt{SM-chunks} of the required tensors are already cached in memory.
We sort Type-I and Type-II tasks independently in non-increasing order of their corresponding expert execution time $p_{n(j)}$, yielding two ordered sequences $\sigma_{\mathrm{I}}$ and $\sigma_{\mathrm{II}}$, respectively.\\
\noindent
\textbf{Block Construction.}
Building on the insight that \texttt{SM-chunk} loading dominates the critical path of \emph{Type-I} tasks, the scheduler creates blocks by grouping \emph{Type-I} tasks and \emph{Type-II} tasks to overlap \texttt{SM-chunk} loading and decompression.
Specifically, the scheduler iteratively selects the first unscheduled task in $\sigma_{\mathrm{I}}$ as the \emph{base} of a new block.
Within each block, \texttt{E-chunks} are loaded before \texttt{SM-chunks}, and the I/O order among the same type of chunks follows the scheduling order of their corresponding tasks in the block.
The scheduler then incrementally inserts tasks from the head of $\sigma_{\mathrm{II}}$ into the block at positions that do not introduce additional idle time on the decompression threads.
If no such position exists, the task is placed after all \emph{Type-II} tasks (or \emph{Type-I} tasks, if no \emph{Type-II} task exists) with more token requests.
The insertion process terminates when either all Type-II tasks have been considered, or when the system becomes compute-bound, i.e., when the completion time of the $l$-th ( $1\leq l\leq \min\{L, K\}$) earliest decompression thread lags behind the I/O thread by at least $\frac{l \cdot \rho}{K} \cdot u$.
Beyond this point, additional Type-II tasks no longer contribute to hiding I/O latency and the resulting sequence of tasks forms a block.
The scheduler repeatedly applies this procedure to construct subsequent blocks until all tasks are assigned.
Blocks are executed sequentially, while tasks within each block follow the constructed priority order.
We provide the full pseudocode in Appendix \ref{sec:code_scheduler} and prove that the resulting schedule achieves a constant-bounded approximation to the optimal makespan in Appendix \ref{sec:proof_scheduler}.

\begin{theorem}
\label{thm:approximation_ratio}
Let \texttt{ALG} be the makespan achieved by \textsc{ZipMoE}'s scheduler and \texttt{OPT} be the optimal value. It holds that:
\begin{equation}
\label{eq:approximation_ratio}
\texttt{ALG}\leq\left (3-\frac{1}{L}\right )\cdot\texttt{OPT},
\end{equation}
where $L$ is the number of decompression threads.
\end{theorem}

\noindent
\textbf{Memory-Coalesced Tensor Recovery Kernel.}
Reassembling exponent bits and sign-mantissa bits into BF16 tensors is inherently memory-bound, where the computation consists only of lightweight bit manipulations, while the kernel must stream a large volume of tensor elements from global memory.
We optimize this process via a vectorized GPU kernel that aggregates data access to ensure memory coalescing. Specifically, each CUDA thread fetches contiguous segments of the loaded \texttt{SM-chunks} and the decompressed \texttt{E-chunks} via vectorized load instructions.
This design allows bit-level reconstruction to be performed entirely within registers and writes the recovered tensors back using vectorized stores.
As a result, the kernel minimizes instruction overhead, saturates DRAM bandwidth, and achieves high-throughput tensor recovery that effectively overlaps with the asynchronous CPU decompression pipeline.

\subsection{Compression-Aware Cache Management}
\label{sec:cache_management}

To enable fine-grained memory control, \textsc{ZipMoE} partitions the total available memory into a hierarchy of cache pools: the $\mathcal{F}$ pool for full tensors, the $\mathcal{C}$ pool for compressed tensors (containing both \texttt{E-chunks} and \texttt{SM-chunks}), the $\mathcal{S}$ pool dedicated to \texttt{SM-chunks}, and the $\mathcal{E}$ pool dedicated to \texttt{E-chunks}. 
These pools offer different latency--memory trade-offs and collectively enable flexible cache management under a fixed memory budget.\\
\noindent
\textbf{Rank-Based Workload Modeling.}
In MoE workloads, the specific identities of frequently activated experts often fluctuate across different prompts.
To capture the inherent workload skewness while remaining agnostic to expert identities, we introduce a \emph{rank-based abstraction}.
Specifically, we aggregate historical expert activation counts for a given batch size to derive a rank-based marginal inclusion probability list $(f_r)_{r\in\mathcal{N}}$.
This list characterizes the stationary distribution of expert activations with respect to their popularity ranks, independent of the particular expert IDs occupying those ranks at runtime.\\
\noindent
\textbf{Pool Dispatching.}
The system maintains a \emph{runtime} frequency list to record the actual activation counts of each expert.
We define an ordered cache hierarchy $\mathcal{F} \prec \mathcal{C} \prec \mathcal{S} \prec \mathcal{E}$.
An expert is dispatched to the first cache pool $i$ in the hierarchy for which its observed rank $r$ satisfies
$r < \tau_i = \sum_{j: j \preceq i} S_j + \delta$, where $S_j$ denotes the capacity of pool $j$, and $\delta$ is a \emph{tolerance margin} introduced to absorb noise in runtime statistics that may deviate from the long-term ranking implied by the workload model.
In cases where the assigned cache pool overflows, experts with the lowest activation frequency are evicted.
If an expert’s rank exceeds all thresholds, it is treated as rarely activated and is evicted immediately after execution.\\
\noindent
\textbf{Hierarchical Cache Planning.}
We determine the optimal cache partition by minimizing the expected makespan of sparse MoE layers.
To this end, we analytically model the cache behavior of feasible partitions using the rank-based workload abstraction and the defined cache hierarchy.
Aligning with the runtime dispatching logic, we map the cache pools onto the \emph{expert frequency ranks} sequentially.
Consequently, each pool $p$ is assigned to a \emph{contiguous rank interval} $[u_p, v_p]$ with a size equal to its capacity $S_p$.
Expert ranks falling outside all allocated cache pools (i.e., the tail of the rank list) are assigned to a virtual \emph{Miss Pool}, denoted by $\mathcal{M}$.
We model the expert activation process as a series of independent Bernoulli trials.
Specifically, for a pool $p$ (including $\mathcal{M}$), an expert at rank $r \in [u_p, v_p]$ is selected with probability $q_r$, which is derived from the marginal probability $f_r$ (details in Appendix \ref{sec:proof_entropy}).
Under this formulation, the number of cache hits within any pool follows a \emph{Poisson binomial distribution}.
We employ dynamic programming (DP) to compute the probability of observing exactly $h_p$ hits within the specific rank interval of pool $p$.
Let $\mathrm{DP}[i][j][p]$ denote the probability of observing $j$ hits among the first $i$ experts within the interval assigned to pool $p$.
The DP table transition is given by:
$
\mathrm{DP}[i][j][p] = \mathrm{DP}[i-1][j][p] (1 - q_{u_p + i}) + \mathrm{DP}[i-1][j-1][p] q_{u_p + i}
$, where $q_{u_p + i}$ denotes the probability of selecting the $i$-th expert in the pool's assigned interval.
The probability of observing exactly $h_p$ hits in pool $p$ is thus given by $\Phi_p(h_p) = \mathrm{DP}[S_p][h_p][p]$.
Finally, since exactly $k$ experts are selected per token layer-wise, we derive the joint probability of a cache hit pattern $\mathbf{h} = (h_{\mathcal{F}}, h_{\mathcal{C}}, h_{\mathcal{S}}, h_{\mathcal{E}})$ as a conditional probability:
$
    \mathbb{P}\left(\mathbf{h} \mid \sum_{p \in \Lambda \cup \{\mathcal{M}\}} h_p = k\right)
    =
    \frac{
    \Phi_{\mathcal{M}}(k_{\mathrm{rem}}) 
    }{
    \Phi_{\mathcal{N}}(k)
    }\cdot \prod_{p \in\Lambda} \Phi_p(h_p)
$
, where $\Lambda \in \left \{\mathcal{F}, \mathcal{C}, \mathcal{S}, \mathcal{E} \right \}$ is the set of activated cache pools, which can be selected flexibly by users based on specific hardware and OS conditions.
$k_{\mathrm{rem}} = k - \sum_{p \in \Lambda} h_p$ is the number of hits falling into the \emph{Miss Pool},
$\Phi_{\mathcal{M}}$ represents the probability distribution of the tail experts,
and $\Phi_{\mathcal{N}}(k)$ is the probability of selecting exactly $k$ experts from the entire rank list $\mathcal{N}$, computed from the global DP table.
We prove in Appendix \ref{sec:proof_entropy} that this procedure yields a maximum entropy distribution, ensuring the solution is \emph{Bayes robust} \cite{AS04}.

\begin{theorem}
\label{thm:max_ent}
Among all probability distributions over $k$-sized expert subsets that are consistent with the observed individual expert selection counts, the distribution produced  by the DP procedure achieves maximum entropy.
\end{theorem}

The planning algorithm then enumerates feasible cache partition configurations via grid search. 
For each configuration, we compute the expected makespan by aggregating the latency costs of all feasible cache hit and miss combinations weighted by their corresponding probabilities derived above. 
The configuration that minimizes the expected makespan is selected for runtime cache partitioning. 
The complete planning algorithm is detailed in Appendix \ref{sec:code_cache}.

\section{Implementation}

We prototype \textsc{ZipMoE} on top of the HuggingFace Transformers \cite{Transformers} framework for inference, and integrate the \emph{lz4} and \emph{zstd} libraries as lossless compression backends.
The system frontend, profiler, and cache planning components are implemented in 2.6K lines of Python code, while the execution engine, scheduler, and memory management system comprise 8K lines of C++/CUDA code.
\textsc{ZipMoE} pre-allocates cache pools as contiguous memory regions to reduce memory fragmentation.
To benefit from UMA, we adopt a zero-copy paradigm that reads \texttt{SM-chunks} directly into host-pinned memory registered for direct GPU access, avoiding redundant data transfers.
Motivated by operating system (OS) considerations, the prototype consolidates each \texttt{E-chunk} read with its subsequent decompression into a single operation to better leverage the page cache.
This implementation preserves the design principles in \S\ref{sec:design} while simplifying runtime overhead and improving affinity with the OS.

\section{Evaluation}

\begin{figure}
\centering
\includegraphics[scale=0.42]{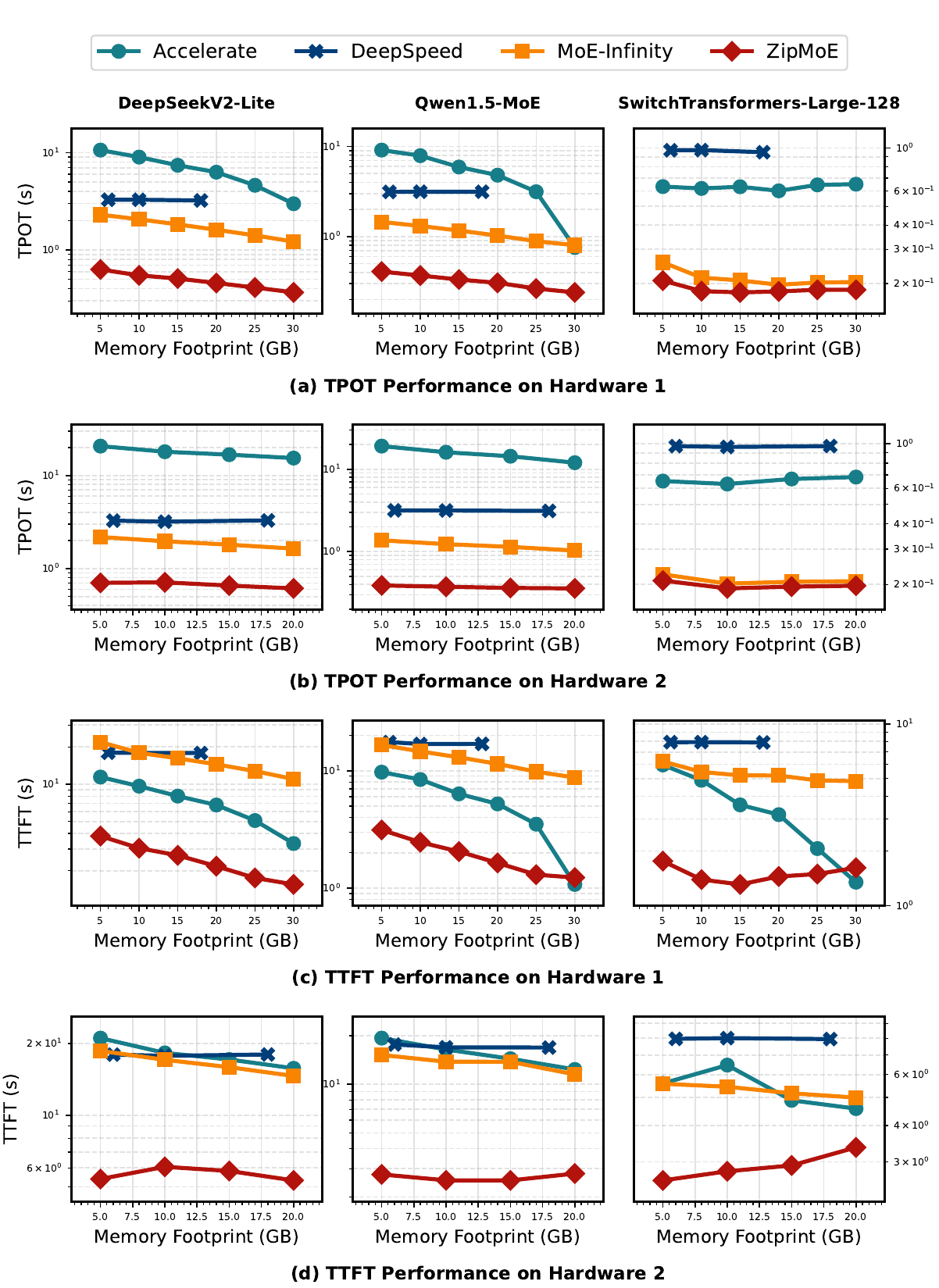}
\caption{Comparison of TPOT and TTFT performances on diverse MoE models across different memory budgets.}
\label{fig:Latency}
\end{figure}

\begin{figure}
\centering
\includegraphics[scale=0.45]{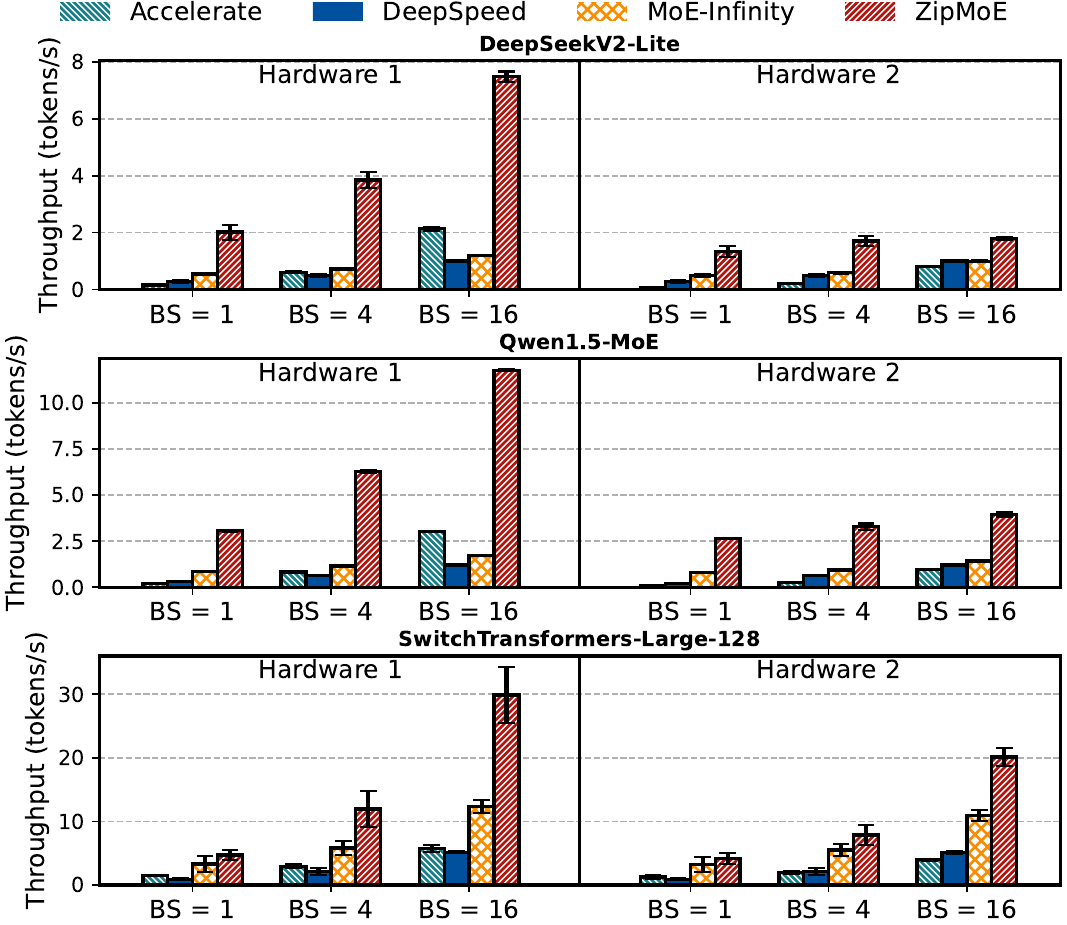}
\caption{
Comparison of system throughput on diverse MoE models under different batch sizes.
BS denotes batch size.
}
\label{fig:TPS}
\end{figure}

\begin{figure*}[t]
    \centering
    \includegraphics[width=0.98\textwidth]{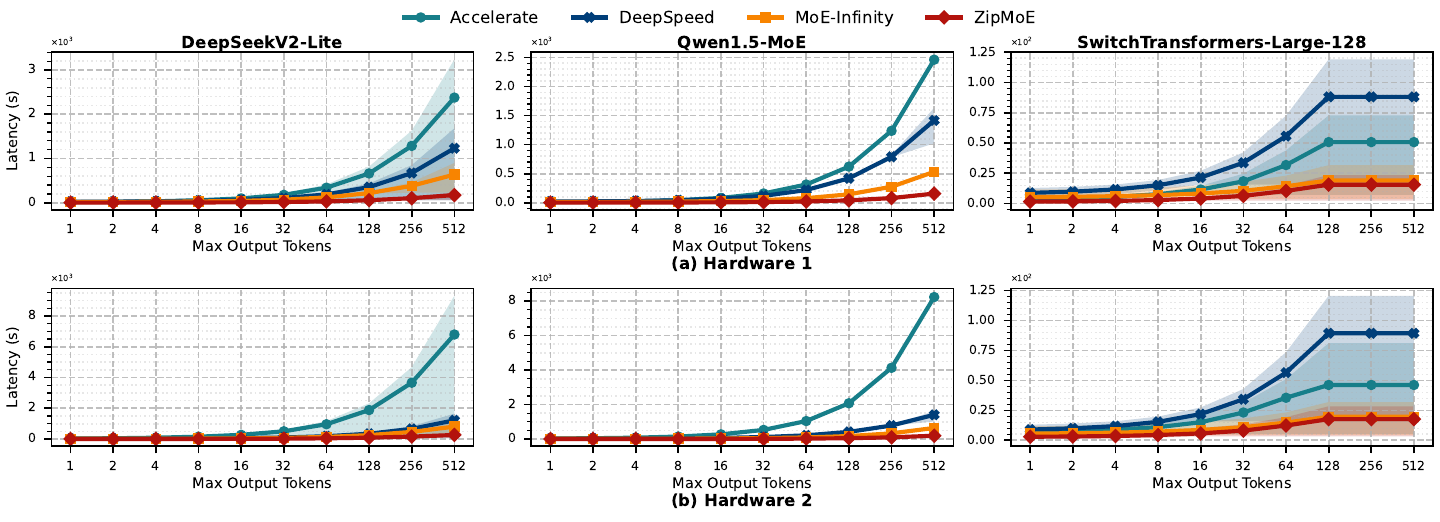}
    \caption{Comparison of end-to-end latency on diverse MoE models across different output lengths.}
    \label{fig:E2E}
\end{figure*}

\begin{figure}
\centering
\includegraphics[scale=0.38]{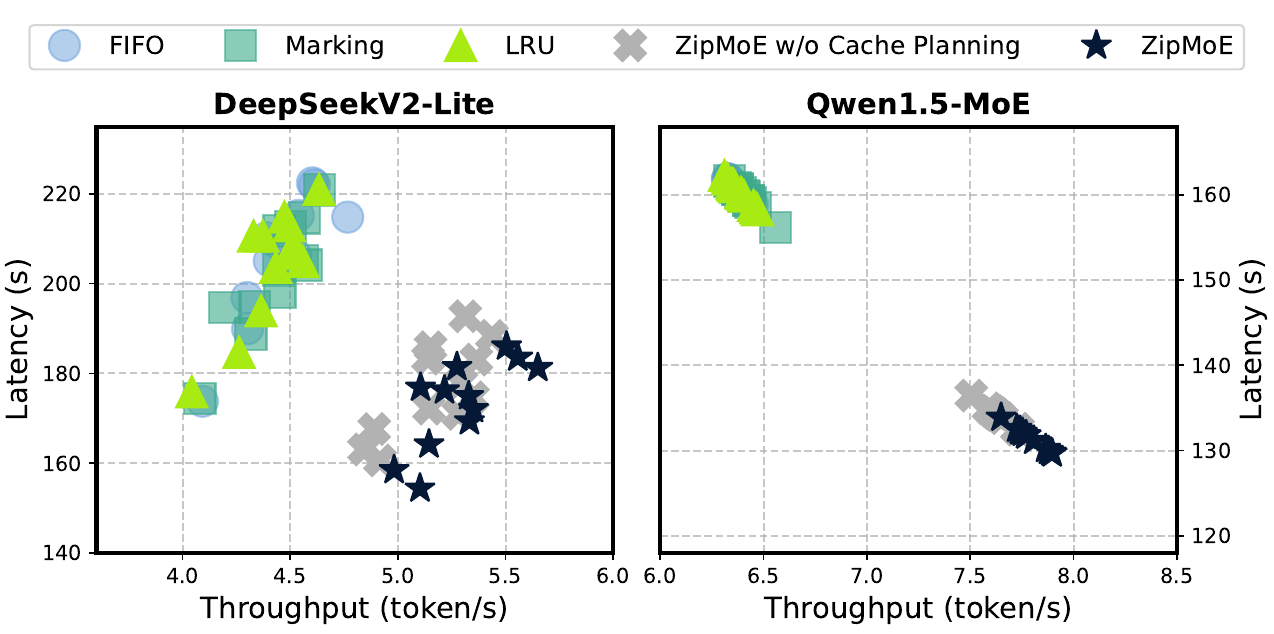}
\caption{Impact of cache management strategies on the trade-off between latency and throughput.}
\label{fig:Ablation}
\end{figure}

\noindent
\textbf{Models and Datasets.}
We evaluate \textsc{ZipMoE} on two decoder-only models, DeepSeekV2-Lite \cite{DeepSeek} and Qwen1.5-MoE \cite{Qwen}, as well as an encoder-decoder model, SwitchTransformers-Large-128 \cite{Switch}.
 All models are obtained from Hugging Face without modification. We randomly sample prompts from the ShareGPT \cite{ShareGPT} dataset and align the inputs across all baseline systems to ensure fair evaluation.\\
\noindent
\textbf{Hardware.}
We adopt the Jetson AGX Orin 64GB (Hardware 1) and 32GB (Hardware 2) as our edge computing testbeds.
Both devices operate on Jetpack 6.2.1 with Ubuntu 22.04, and are equipped with a Samsung 970 EVO SSD that provides a disk read speed of 3.5 GB/s.\\
\noindent
\textbf{Baselines.}
We compare \textsc{ZipMoE} with the following state-of-the-art LLM serving systems: 
(1) \emph{MoE-Infinity} \cite{MoEInfinity}: A high-performance MoE serving system that exploits sparsity-aware expert caching and prefetching. 
(2) \emph{DeepSpeed} \cite{DeepSpeed}: A leading inference engine that supports model offloading and parameter partitioning. We adopt DeepSpeed ZeRO-3 for SSD offloading during MoE inference. 
(3) \emph{Accelerate} \cite{Accelerate}: A widely-used library that provides simple APIs for model offloading and distributed inference. 
All systems are configured to operate under their default SSD offloading modes with aligned DRAM memory footprints for fair comparison.

\subsection{Experimental Results}

\begin{figure}
\centering
\includegraphics[scale=0.38]{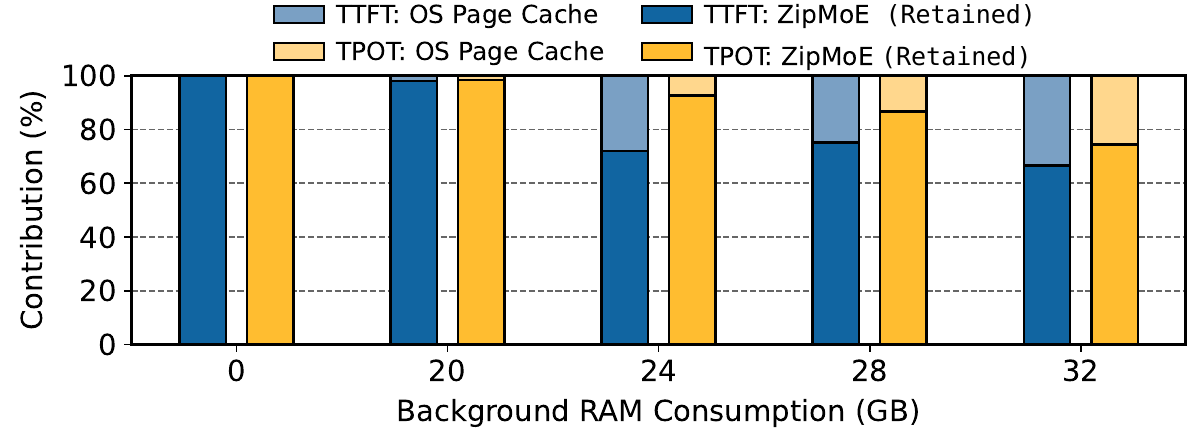}
\caption{Breakdown of contributions under different background RAM pressures.}
\label{fig:OSPageCache}
\end{figure}

\noindent
\textbf{Real-Time Responsiveness.}
Fig. \ref{fig:Latency} shows the real-time responsiveness measured by Time-Per-Output-Token (TPOT) and Time-To-First-Token (TTFT). 
For decoder-only models, we observe that \textsc{ZipMoE} substantially improves performance on both hardware environments, achieving $62.65\%$-$97.97\%$ TPOT reduction and $53.25\%$-$87.90\%$ TTFT reduction in scenarios where offloading is mandatory.
For encoder-decoder models, although the advantage diminishes due to highly skewed expert activation and a less I/O-intensive structure, \textsc{ZipMoE} still yields a $4.99\%$-$81.24\%$ TPOT reduction and up to an $83.45\%$ TTFT reduction.
These advantages are enabled not only by the sophisticated caching-scheduling co-design, but also by the low-overhead implementation and its effective utilization of the page cache during fixed-memory experiments, which improves both memory and I/O efficiency.
The results highlight the \textsc{ZipMoE}'s superiority in real-time and interactive tasks.\\
\textbf{System Throughput.}
Fig. \ref{fig:TPS} compares the throughput of different systems in batch inference settings with batch sizes of $1$, $4$, and $16$.
The memory budget for \textsc{ZipMoE}, \emph{MoE-Infinity}, and \emph{Accelerate} is fixed at $20$GB and $10$GB on Hardware 1 and Hardware 2, respectively.
For \emph{DeepSpeed}, we observe extreme discontinuity between its runtime memory footprint and configuration, therefore memory budget control is coarse-grained.
We fix its runtime memory budget to $18$GB and $10$GB on Hardware 1 and Hardware 2, respectively.
Note that since \emph{DeepSpeed} employs a sliding-window offloading pipeline, its performance remains agnostic to the specific memory footprint provided the budget is below the model size (see Fig. \ref{fig:Latency}).
It can be observed that \textsc{ZipMoE} consistently surpasses all baselines, achieving $1.79\times$-$42.49\times$ improvement for decoder-only models and $1.31\times$-$5.82\times$ improvement for encoder-decoder models.
These advantages are attributed to \textsc{ZipMoE}'s scheduler, which is specifically designed for I/O efficiency via decompression parallelization. As more experts are activated per layer in batch inference, the benefits of \textsc{ZipMoE}'s scheduler for parallelization are amplified, thus significantly reducing I/O blocking.
Overall, we conclude that \textsc{ZipMoE} also excels in batch inference for offline background tasks.\\
\textbf{End-To-End Latency.}
Fig. \ref{fig:E2E} presents the end-to-end latency results under different output token limits, where the lines represent the average values and the shaded areas cover between the top and bottom $10\%$ values.
Note that sublinear trends may appear in the figure, as in certain cases the number of output tokens do not actually reach the limit.
It can be observed that \textsc{ZipMoE} is consistently superior to all baselines across all maximum output token numbers, accelerating end-to-end inference speed by $3.03\times$-$42.49\times$ for decoder-only models and $1.11\times$-$5.64\times$ for encoder-decoder models.
These results confirm \textsc{ZipMoE}'s versatility in providing consistent inference acceleration across diverse MoE architectures and hardware environments.\\
\textbf{Ablation Study.}
Fig. \ref{fig:Ablation} illustrates the impact of the cache management module on system performance in terms of the latency-throughput trade-off.
We successively isolate the hierarchical cache planning algorithm and replace the eviction strategy of \emph{ZipMoE} with First-In-First-Out (FIFO), Marking\cite{Marking}, and Least Recently Used (LRU).
All experiments are conducted under 10GB memory footprint and batch size=16. 
It can be observed that while all baseline caching algorithms perform similarly, \textsc{ZipMoE}'s built-in caching algorithm outperforms all of them.
Furthermore, with cache planning that optimizes the cache pools prior to inference, \textsc{ZipMoE}'s performance is further boosted.
To further quantify the contribution of each component, we report a detailed breakdown of performance gains in Tab. \ref{tab:ablation}.
Here, \emph{Avg. basic} represents the average performance of LRU, FIFO, and Marking, \emph{C} denotes \textsc{ZipMoE}’s heterogeneous cache pool (without planning), and \emph{P} denotes cache pool planning.
The results indicate that the core system accounts for the majority of the throughput improvement, constituting approximately 76\% of the total gain.
This leap is primarily driven by the paradigm shift enabled by \textsc{ZipMoE}'s lossless decompression engine and CPU-parallel decompression scheduling.
It is worth noting that these gains stem not only from the decompression engine itself, but also from optimized system-level behaviors, including improved utilization of the OS page cache, more efficient GPU memory management, and refined low-level optimizations in the inference framework.
Our proposed cache management approach contributes the remaining approximately 24\% of the throughput improvement.
Compared with using heterogeneous cache pools alone, the cache planning algorithm provides a moderate performance gain while achieving lower latency and higher throughput in a Pareto-optimal manner.\\
\textbf{Isolating OS Page Cache.}
\textsc{ZipMoE} elastically utilizes the OS page cache whenever additional RAM is available and not occupied by background processes from co-located applications.
To investigate the performance contribution of OS page cache utilization, we artificially inject background memory consumption during inference to progressively reduce the available RAM for page caching.
The results are presented in Tab.~\ref{tab:page_cache_isolation} and Fig.~\ref{fig:OSPageCache}.
Since the OS page cache is elastic, increasing background memory pressure effectively reduces the cacheable memory space, forcing the system to rely primarily on \textsc{ZipMoE}'s internal caching and decompression mechanisms.
The results show that even when background RAM consumption reaches 32\,GB, leaving negligible space for the OS page cache, \textsc{ZipMoE} still achieves a 56.64\% reduction in TTFT and a 53.32\% reduction in TPOT compared with the baseline, retaining approximately 66.7\%--74.4\% of its original performance advantage.
The results confirm that the majority of the performance gains stem from the core \textsc{ZipMoE} system engine, while the OS page cache provides an opportunistic performance enhancement when additional memory resources are available.

\begin{table}[t]
\centering
\caption{
Quantitative breakdown of \textsc{ZipMoE}’s performance gains across different cache management strategies.
}
\label{tab:ablation}

\begin{subtable}[t]{\columnwidth}
\centering
\caption{Model: DeepSeekV2-Lite 16B}
\resizebox{\columnwidth}{!}{
\label{tab:deepseek_breakdown}
\begin{tabular}{lcccc}
\toprule
\textbf{Method} & \makecell[c]{\textbf{Throughput} \\ (tokens/s)} & \makecell[c]{$\Delta$\textbf{Throughput} \\ (tokens/s)} & \textbf{E2E} (s) & \textbf{$\Delta$E2E} (s) \\ 
\toprule
Baseline            & 1.60 & -     & 585.96 & -       \\ 
ZipMoE (avg. basic) & 4.43 & +2.83 & 204.05 & -381.91 \\ 
ZipMoE (+C)         & 5.18 & +0.75 & 176.68 & -27.37  \\ 
ZipMoE (+C+P)       & 5.30 & +0.12 & 173.23 & -3.45   \\ 
\midrule
Total gain          & +3.70 & -    & -412.73 & -      \\ 
\bottomrule
\end{tabular}
}
\end{subtable}

\vspace{1em}

\begin{subtable}[t]{\columnwidth}
\centering
\caption{Model: Qwen1.5-MoE 14B}
\resizebox{\columnwidth}{!}{
\label{tab:qwen_breakdown}
\begin{tabular}{lcccc}
\toprule
\textbf{Method} & \makecell[c]{\textbf{Throughput} \\ (tokens/s)} & \makecell[c]{$\Delta$\textbf{Throughput} \\ (tokens/s)} & \textbf{E2E} (s) & \textbf{$\Delta$E2E} (s) \\ 
\toprule
Baseline            & 1.99 & -     & 515.12 & -       \\ 
ZipMoE (avg. basic) & 6.39 & +4.40 & 160.33 & -354.79 \\ 
ZipMoE (+C)         & 7.64 & +1.25 & 134.10 & -26.23  \\ 
ZipMoE (+C+P)       & 7.79 & +0.15 & 131.47 & -2.63   \\ 
\midrule
Total gain          & +5.80 & -    & -383.65 & -      \\ 
\bottomrule
\end{tabular}
}
\end{subtable}

\end{table}

%
%

\section{Discussion}

\textbf{Hardware Generalizability.}
Although the current evaluation of \textsc{ZipMoE} is conducted on NVIDIA Jetson AGX Orin, the overall system design is intentionally hardware-agnostic and applicable to any shared-memory platforms.
For systems in which the CPU and GPU (or other accelerators) maintain dedicated physical memory, our approach can be extended by treating CPU memory as an additional cache tier and introducing asynchronous host-to-device transfers to support CPU-GPU hybrid offloading and inference.
We consider extending \textsc{ZipMoE} to such heterogeneous-memory systems an important direction for future work.\\
\textbf{Other Compressors.}
Our current implementation supports multiple compression backends, including LZ4, LZ4HC, and ZSTD.
Since the effectiveness of a compressor depends on both hardware characteristics and implementation efficiency, our codebase provides APIs that allow future integration of customized or hardware-specific compressors for further performance optimization.

\newcolumntype{Y}{>{\raggedright\arraybackslash}X}
\begin{table}[t]
\centering
\caption{
Isolation of page cache gain. 
\emph{Reduction} denotes the percentage of reduced latency.
\emph{Retention} denotes the percentage of performance advantage over the baseline (MoE-Infinity) that is preserved as the page cache is reduced.
}
\label{tab:page_cache_isolation}
\setlength{\tabcolsep}{4pt} %
\resizebox{\columnwidth}{!}{ 
\scriptsize
\renewcommand{\arraystretch}{1.3} 
\begin{tabularx}{\linewidth}{>{\raggedright\arraybackslash}p{4em}|YYY|YYY} 
\Xhline{1pt}
\multirow{2}{*}{\makecell[l]{\textbf{Occupied} \\ \textbf{RAM}}} & \multicolumn{3}{c|}{\textbf{Time-To-First-Token}} & \multicolumn{3}{c}{\textbf{Time-Per-Output-Token}} \\
\cline{2-7} 
 & \textbf{Value} & \textbf{Reduction} & \textbf{Retention} & \textbf{Value} & \textbf{Reduction} & \textbf{Retention} \\ 
\Xhline{1pt} 
0 GB  & 2.18 s & 84.94\% & 100.0\% & 0.46 s & 71.64\% & 100.0\% \\ \hline
20 GB & 2.43 s & 83.21\% & 98.0\%  & 0.48 s & 70.46\% & 98.4\%  \\ \hline
24 GB & 5.62 s & 61.17\% & 72.0\%  & 0.54 s & 66.41\% & 92.7\%  \\ \hline
28 GB & 5.22 s & 63.90\% & 75.2\%  & 0.61 s & 62.18\% & 86.8\%  \\ \hline
32 GB & 6.27 s & 56.64\% & 66.7\%  & 0.75 s & 53.32\% & 74.4\%  \\  
\Xhline{1pt} 
\textbf{Baseline} & 14.46 s & - & - & 1.61 s & - & - \\ 
\Xhline{1pt} 
\end{tabularx}
}
\end{table}

\section{Conclusion}
We proposed \textsc{ZipMoE}, an efficient MoE inference system tailored for mobile and edge computing platforms.
\textsc{ZipMoE} improves on-device memory and I/O efficiency through fine-grained orchestration of sparse layer execution via a caching-scheduling co-design.
Our testbed evaluation of \textsc{ZipMoE} on diverse edge platforms demonstrates significant improvements in both inference latency and system throughput while preserving identical model behavior.

%

\section*{Acknowledgements}
The authors would like to thank the anonymous reviewers, whose invaluable comments helped improve the presentation of this paper substantially.
This work was partially supported by the Jiangsu Science Foundation Leading-edge Technology Program (BK20232003).

\section*{Impact Statement}
This paper presents work whose goal is to advance the field of machine learning. 
There are many potential societal consequences of our work, none of which we feel must be specifically highlighted here.

\bibliography{ZipMoE_reference}
\bibliographystyle{icml2026}

\newpage
\appendix
\onecolumn
\section{\textsc{ZipMoE}'s Scheduler}
\label{sec:code_scheduler}

In this section, we describe \textsc{ZipMoE}'s scheduler in detail.
Using offline-profiled decompression and I/O delays, the scheduler constructs blocks via \emph{Algorithm} \ref{alg:block_priority}.
Then the compute workers and the I/O thread fetch the first available operations in a work-conserving manner, following the priority order of the constructed blocks.

\begin{algorithm}[h]
\caption{Cache-Affinity Scheduling}
\label{alg:block_priority}
\begin{algorithmic}[1]

\State \textbf{Input:} DAG Information in request set $\mathcal{Q}$.
\State \textbf{Output:} An ordered set $\mathcal{B}$ containing the constructed blocks.

\State Partition $\mathcal{Q}$ into \emph{Type-I} tasks $\mathcal{Q}_\text{I}$ and  \emph{Type-II} tasks $\mathcal{Q}_\text{II}$
\State $\sigma_\text{I} \gets$ Sort tasks $i \in \mathcal{Q}_\text{I}$ in non-increasing order of $p_n(i)$, and group tasks from the same expert consecutively.
\State $\sigma_\text{II} \gets$ Sort tasks $j \in \mathcal{Q}_\text{II}$ in non-increasing order of $p_n(j)$, and group tasks from the same expert consecutively.
\State Initialize $\mathcal{B} \gets \emptyset$

\While{$\sigma_\text{I}$ is \textbf{not} empty}
    \State Create ordered list $U \gets (\sigma_{II}, \sigma_{I})$
    \State Create new block $B \gets \emptyset$
    \State Append the task in $\sigma_{I}$ with the highest priority to $B$

    \While{$B$ is \textbf{not} compute-dominant}
        \State $j^* \gets$ task in $U$ with the highest priority
	\State Attempt to insert $j^*$ at the earliest position in $B$ that introduces no additional idle period on any thread
	
	\If{no such position exists}
	    \If{there exists a type-II job}
	        \State Append $j^*$ to the tail of any \emph{Type-II} job $i^*$ with $p_{i^*} \ge p_{j^*}$
	    \Else
	        \State Append $j^*$ to the tail of some \emph{Type-I} job $i^*$ with $p_{i^*} \ge p_{j^*}$
	    \EndIf
	\EndIf
        
        \State Remove $\{j^*\}$: $U \gets U \setminus \{j^*\}$
        \If{$U = \emptyset$}
            \State \textbf{break}
        \EndIf
    \EndWhile

    \State Append $B$ to the back of $\mathcal{B}$: $\mathcal{B} \gets (\mathcal{B}, B)$
\EndWhile

\State \textbf{return} Ordered set $\mathcal{B}=(B_1, B_2, \cdots, B_M)$

\end{algorithmic}
\end{algorithm}

In \emph{Algorithm} \ref{alg:block_priority}, the compute-bound property is formally defined as follows.

\begin{definition}[Compute-Bound]
\label{def:compute_bound}
Let $\mathcal{L}$ be the set of compute threads with $|\mathcal{L}|=L$.
For each block $B_i, i\in [M]$, let $F_{I/O}(B_i)$ be the completion time of its I/O thread, and let $F_c^{(l)}(B_i)$ be the completion time of its $l$-th compute thread.
A block is compute-dominant if it holds for all $l\in\mathcal{L}$ that:
\begin{equation}
\label{eq:compute_dominate}
F_c^{(l)}(B_i)-F_{I/O}(B_i)\geq k\cdot\frac{\rho}{K}\cdot u, 0\leq k\leq \min\{L,K\}.
\end{equation}
\end{definition}

\newpage
\section{Theoretical Analysis of \textsc{ZipMoE}'s Scheduling Policy}
\label{sec:proof_scheduler}

In this section, we prove the approximation ratio of our proposed scheduling algorithm.
Our objective is to minimize the completion time of each sparse layer. 
This can be modeled using the concept of \emph{makespan}, as formally defined in classic scheduling literature \cite{DAG1, DAG2, DAG3}.
We start our theoretical analysis by defining some necessary notions.

\begin{definition}[Charge]
\label{def:charge}
For any schedule $A$, let $\delta_{v,j}(A)$ denote the idle time interval between the end of the last active operation on the thread processing chunk $v$ of job $j$ and the start of its decompression.
We define $\tilde{c}_{v,j}(A):=c_{v,j}+\delta_{v,j}(A)$ to be the charged time of operation $v$ of job $j$.
Further, the total charged time of any algorithm $A$ is:
\begin{equation}
\label{eq:charge}
\Delta(A) = \sum_{j\in\mathcal{Q}}\sum_{v\in D_j}\delta_{v,j}(A),
\end{equation}
where $D_j$ is the set of decompression operations of job $j$.
\end{definition}

\emph{Remark.} Intuitively, the charged time of any decompression operation is the idle time before execution on its compute thread, which is caused by I/O blocking of its required (and possibly other tasks with higher priorities within the block) \texttt{E-chunks}.

\begin{definition}[Critical-Path]
\label{def:critical_path}
Let $\tau_j\in\{\texttt{M}, \texttt{C}, \texttt{S}, \texttt{E}\}$ be the compression state (expert-type) of task $j$, representing missed expert, compressed expert, \texttt{SM-expert}, and \texttt{E-expert}, respectively.
For any task $j\in\mathcal{Q}$, the critical path $z_j$ is the least time required to finish all the workloads following the precedence constraints, that is
\begin{equation}
\label{eq:charge}
z_j = \rho u \cdot\mathbbm{1}_{\tau_j\in\left \{\texttt{M}, \texttt{S} \right \}}+\max\left \{\frac{K\cdot c }{\min\left \{ K,L\right \}},u\cdot\mathbbm{1}_{\tau_j\in\left \{\texttt{M}, \texttt{E} \right \}} \right \}+p_j.
\end{equation}
\end{definition}


We can first provide the lower-bound for the optimal solution.

\begin{lemma}
\label{lem:OPT}
Define $v_j$ as the I/O workload of task $j\in\mathcal{Q}$, i.e., $v_j=(1+\rho)u\mathbbm{1}_{\tau_j\in\texttt{M}}+u\mathbbm{1}_{\tau_j\in\texttt{E}}+\rho u\mathbbm{1}_{\tau_j\in\texttt{S}}$.
Let $I:=\sum_{j\in\mathcal{Q}}v_j$ be the total I/O workload, $C:=\sum_{j\in\mathcal{Q}}K\cdot c$ be the total compute workload, $P:=\sum_{j\in\mathcal{Q}}\frac{p_{n(j)}}{|\{i:n(i)=n(j)\}|}$ be the total token workload, and $Z:=\max_{j\in \mathcal{Q}}\left \{ z_j\right \}$ be the longest single-job critical path.
We have
\begin{equation}
\label{eq:lemma1}
\texttt{OPT}\geq \max\left\{ I, \frac{C}{L}, P, Z \right\}.
\end{equation}
\end{lemma}

\begin{proof}
The lemma is trivial, since each of the lower bounds is in the critical path of the set of requests, any of them must be smaller than $\texttt{OPT}$.
\end{proof}

\begin{corollary}
\label{cor:OPT}
We also have $\texttt{OPT}\geq \frac{C+\Delta(\texttt{OPT})}{L}$.
\end{corollary}

Next, we prove that the total charged time in our proposed algorithm (\texttt{ALG}) lower-bounds the optimal solution (\texttt{OPT}).

\begin{lemma}
The total charge of the optimal solution cannot be lower than Algorithm \ref{alg:block_priority}. It holds that:
\begin{equation}
\label{eq:lemma2}
\Delta(\texttt{ALG})\leq\Delta(\texttt{OPT}).
\end{equation}
\end{lemma}

\begin{proof}
We introduce an auxiliary schedule $A'$. In $A'$, the I/O thread prioritizes loading the compressed \texttt{E-chunks} required for decompression above all else, and the compute thread pool operates in a work-conserving manner, executing any ready decompression task immediately.
Note that in our setting, decompression bubbles (idle periods on worker threads) are only caused by the unavailability of \texttt{E-chunks} due to I/O latency.
Since $A'$ minimizes the waiting time for these dependencies by prioritizing their I/O, it represents a theoretical lower bound on idle time. Thus, $\Delta(A')\leq\Delta(\texttt{OPT})$.

Now, we compare $\Delta(\texttt{ALG})$ with $\Delta(A')$.
Let $\mathcal{B} = \{B_1, B_2, \dots, B_M\}$ be the ordered set of blocks constructed by \texttt{ALG}.
According to the block Construction policy, a block is finalized and pushed to $\mathcal{B}$ only when it becomes \emph{compute-bound} (i.e., the estimated decompression completion lags sufficiently behind I/O) or when no tasks remain in $\sigma_{\mathrm{I}} \cup \sigma_{\mathrm{II}}$.

Consider the final block $B_M$. We distinguish two cases:

\emph{CASE I: There is no bubble in block $B_M$.}

In this case, we claim that $B_m$ must be compute-dominant for every $m<M$.
Since if it is not the case, there has to be an $m<M$ that is not compute-dominant, and by the design of the algorithm we have to add remaining tasks to this block until either it becomes compute-dominant, or no task remains, which means that $m=M$. Both cases contradict to the claim.
Therefore, since every $B_m, m\in[M-1]$ is compute-dominant, the bubbles (if there are any) can only exist within the blocks.
In that case, it must hold that $c <\rho u $, since otherwise no bubble can be produced as the I/O of $K$ exponent chunks must be able to cover a single compute operation. 
If $c <\rho u $, by Definition \ref{def:compute_bound}, blocks will never be compute-bound, thus we only have $M=1$.
Therefore, it holds that either no bubble exists for all $B_m, m\in[M-1]$, or $M=1$.
In both cases, $\Delta(\texttt{ALG})=0\leq\Delta(\texttt{OPT})$.

\emph{CASE II: There exist some bubbles in block $B_M$.}

In this case, we claim that $M=1$ and $\Delta(\texttt{ALG})\leq \Delta(A)$ for every feasible schedule $A$.
To show that our claim holds, note that when bubbles exist within the block, it can only be that the compute cannot cover the I/O of $K$ exponent chunks: $c<\rho u$. 
In this case, $B_{M-1}$ cannot be compute-bound and by the design of our algorithm (\emph{Definition} \ref{def:compute_bound}), jobs in $B_{M-1}$ must be in the same block with $B_{M}$.
By the same reasoning, we conclude that all jobs fall within the same block, so $M=1$.
When $M=1$, since our I/O thread always loads the exponent bits first and the compute thread works in work-conserving manner.
Thus, $\texttt{ALG}$ has the same bubbles as in $A'$, we have $\Delta(\texttt{ALG}) = \Delta(A') \leq \Delta(A)$ for every feasible schedule $A$.

To conclude, we have $\Delta(\texttt{ALG}) \leq \Delta(A') \leq \Delta(\texttt{OPT})$, which proves the lemma.

\end{proof}

Once a schedule is determined, the start time and completion time of any job in its \emph{Gantt} is fixed.
However, we can move the time intervals of decompression operations across threads provided that the target thread's idle period is enough.
This gives chances to reduce the charged time for each individual operation by manipulating the assigned threads of the decompression.
Let each feasible configuration of decompression thread assignment in the \emph{Gantt} chart corresponds to a charging scheme, we have the next lemma to upper-bound the charged times for each individual operation.

\begin{lemma}
\label{lem:charge_bound}
In \texttt{ALG}, there exists a charging such that $\delta_{v,j}(\texttt{ALG})\leq \rho u$ for every $j\in \mathcal{Q}$ and $v\in D_j$.
\end{lemma}
\begin{proof}
Let the start time of operation $v$ of job $j$ be denoted by $s_{v,j}$.
Since in \texttt{ALG}, the latest $s_{v,j}$ should wait for the completion of the last exponent chunk, which takes $K\cdot\frac{\rho}{K}\cdot u = \rho u$ time.
If  $\delta_{v,j}>0$, there must exist a thread that completes the last decompression before the completion of this I/O.
For this thread, there may exist a preceding compute operation that overlaps with the I/O of the eponent chunks of $j$ by $X_{v,j}>0$ amount of time.
So the charged workload can be constructed by $\rho u-X_{v,j}>0$, which is no larger than $\rho u$.
\end{proof}


\begin{theorem}[Restated]
\label{thm:approximation_ratio}
Let \texttt{ALG} be the makespan achieved by \textsc{ZipMoE}'s scheduler and \texttt{OPT} be the optimal value. It holds that:
\begin{equation}
\label{eq:approximation_ratio}
\texttt{ALG}\leq\left (3-\frac{1}{L}\right )\cdot\texttt{OPT},
\end{equation}
where $L$ is the number of decompression threads.
\end{theorem}

\noindent
\begin{proof}
Suppose \texttt{ALG} divides the jobs into $M$ blocks: $\mathcal{B}=\left \{ B_1, B_2, ..., B_M\right \}$. We define the following quantities.
\begin{itemize}
    \item $F_{I}(k)=\sum_{i=1}^k\sum_{j\in B_i}v_j=:I(B_{1\rightarrow k})$: The completion time of all the I/O operations in blocks $B_1, B_2, ..., B_k$.
    \item $F_{C}(k)$: The completion time of all the compute operations in blocks $B_1, B_2, ..., B_k$. Define its sum compute workload as $C(B_{1\rightarrow k})$.
    \item $\tilde{C}(B_{1\rightarrow k}):=\sum_{i=1}^k\sum_{j\in B_i}\sum_{v\in D_j}\tilde{c}_{v,j}\left ( \texttt{ALG}\right )=C(B_{1\rightarrow k})+\Delta\left ( \texttt{ALG}\right ):$ The total charged compute workload of in blocks $B_1, B_2, ..., B_k$ under algorithm $\texttt{ALG}$.
    \item $P(B_k)$: The sum processing time of all the token executions in block $B_k$. Let $P(B_{k\rightarrow m}):=\sum_{i=k}^mP(B_k)$.
\end{itemize}

For any subset of requests $Q$, the latest ready time for the token executions in the subset is the maximum of the time when all I/O operations are ready and the time when all computations are ready.
Therefore, the completion time of $Q$ can be upper-bounded by:
\begin{equation}
\label{eq:cmax1}
\begin{aligned}
\texttt{ALG}&\leq\max_{1\leq k\leq M}\left \{ \max \left \{ F_{I}(k), F_{C}(k)\right \} + P(B_{k\rightarrow M}) \right \}.
\end{aligned} 
\end{equation}

We first argue that the compute completion time $F_{C}(k)$ is upper-bounded by:
\begin{equation}
\label{eq:fck1}
\begin{aligned}
F_{C}(k)&\leq \texttt{OPT}+\left (1-\frac{1}{L}\right )\max_{j\in B_{1\rightarrow k}, v\in D_j}\tilde{c}_{v,j}.
\end{aligned} 
\end{equation}

To show this, we fix the constant $k$ and consider the last operation $v^*$ finished by $\texttt{ALG}$, let $l^*$ be the threads that processes $v^*$, and let $\mathcal{L}_{j^*}$ be the set of active threads (waiting for the charged time is also considered active) at a time arbitrarily shortly before the start of $v^*$. Let $v\rightarrow l$ denote operation $v$ is assigned to thread $l$.
Note that since our schedule is work-conserving, we have:
\begin{equation}
\label{eq:fck2}
\sum_{j\in\sum_{j\in B_{1\rightarrow k}}}\sum_{v\in D_j: v\rightarrow l^*}\tilde{c}_{v,j}\leq\tilde{c}_{v^*,j}+\sum_{j\in\sum_{j\in B_{1\rightarrow k}}}\sum_{v\in D_j: v\rightarrow l}\tilde{c}_{v,j}, \forall l\in\mathcal{L}_{j^*},
\end{equation}
because otherwise we would have assigned the operation to an earlier ready thread.
Therefore, it holds that:
\begin{equation}
\label{eq:t1}
\begin{aligned}
\tilde{C}(B_{1\rightarrow k})
&=\left (\sum_{j\in B_{1\rightarrow k}}\sum_{v\in D_j:v\rightarrow l_1}\tilde{c}_{v,j}\right )+\left (\sum_{j\in B_{1\rightarrow k}}\sum_{v\in D_j:v\rightarrow l_2}\tilde{c}_{v,j}\right )+\cdots+\left ( \sum_{j\in B_{1\rightarrow k}}\sum_{v\in D_j:v\rightarrow l^*}\tilde{c}_{v,j}\right )-\tilde{c}_{v^*,j}+\tilde{c}_{v^*,j}\\
&\geq\sum_{l\in\mathcal{L}\setminus\mathcal{L}_{j^*}}\sum_{j\in B_{1\rightarrow k}}\sum_{v\in D_j:v\rightarrow l}\tilde{c}_{v,j}+|\mathcal{L}_{j^*}|\left (F_{C}(k)-\tilde{c}_{v^*,j}\right )+\tilde{c}_{v^*,j}\\
&= \sum_{l\in\mathcal{L}\setminus\mathcal{L}_{j^*}}\sum_{j\in B_{1\rightarrow k}}\sum_{v\in D_j:v\rightarrow l}\tilde{c}_{v,j}+|\mathcal{L}_{j^*}|F_{C}(k)-\left (|\mathcal{L}_{j^*}|-1\right )\tilde{c}_{v^*,j}, 
\end{aligned} 
\end{equation}
where the inequality follows from Eq. (\ref{eq:fck2}). Rearranging, we have
\begin{equation}
\label{eq:fck3}
\begin{aligned}
F_{C}(k) &\leq \frac{1}{|\mathcal{L}_{j^*}|}\sum_{l\in\mathcal{L}_{j^*}}\sum_{j\in B_{1\rightarrow k}}\sum_{v\in D_j:v\rightarrow l}\tilde{c}_{v,j} + \left ( 1-\frac{1}{|\mathcal{L}_{j^*}|} \right )\tilde{c}_{v^*,j}\\
&\leq \frac{1}{|\mathcal{L}_{j^*}|}\sum_{l\in\mathcal{L}_{j^*}}\sum_{j\in B_{1\rightarrow k}}\sum_{v\in D_j:v\rightarrow l}\tilde{c}_{v,j} + \left ( 1-\frac{1}{L} \right )\tilde{c}_{v^*,j}, 
\end{aligned} 
\end{equation}
where the second inequality follows from the fact that $|\mathcal{L}_{j^*}|\leq L$.
Now, we are left to show that the first term of Eq. (\ref{eq:fck3}) can be upper-bounded by $\texttt{OPT}$.
To see this, we consider two cases.

\noindent
\emph{Case I: $K\geq L$}.
In this case, $\mathcal{L}_{j^*}=\mathcal{L}$ and $\min\left\{K,L\right \}=L$, since the average completion time of all the threads is a lower-bound of the max-completion time, it follows that:
\begin{equation}
\label{eq:cs1}
\begin{aligned}
\frac{1}{|\mathcal{L}_{j^*}|}\sum_{l\in\mathcal{L}_{j^*}}\sum_{j\in B_{1\rightarrow k}}\sum_{v\in D_j:v\rightarrow l}\tilde{c}_{v,j}&=\frac{1}{L}\tilde{C}(B_{1\rightarrow k})\\
& = \frac{1}{L}\left ( {C}(B_{1\rightarrow k})+\Delta(\texttt{ALG}) \right )\\
&\leq \frac{1}{L}\left ( {C}(B_{1\rightarrow k})+\Delta(\texttt{OPT}) \right )\leq\texttt{OPT},
\end{aligned} 
\end{equation}
where the inequality follows from Lemma \hyperref[result:2]{2} and Corollary \hyperref[co:1]{1}.

\noindent
\emph{Case II: $K< L$}.
In this case, $\min\left\{K,L\right \}=K$. 
If $|\mathcal{L}_{j^*}|=L$, then the case is the same with \emph{Case I} and the same bound holds.
If $|\mathcal{L}_{j^*}|<L$, we point out that this can only be the case when $L>|B_{1\rightarrow k}|\cdot K$, since otherwise the last operation $v^*$ was either assigned to a thread with higher workload, or delayed to create idle time on the thread. Both contradict to the work-conservation property of our algorithm. 
Therefore $L>|B_{1\rightarrow k}|\cdot K$, meaning that the threads are abundant, and we have 
\begin{equation}
\label{eq:kl}
\sum_{l\in\mathcal{L}_{j^*}}\sum_{j\in B_{1\rightarrow k}}\sum_{v\in D_j:v\rightarrow l}\tilde{c}_{v,j} = \texttt{OPT} ,
\end{equation}
which implies Eq. (\ref{eq:fck1}).

Let $X_{b}$ denote the event that all $j\in\mathcal{Q}$ are either \texttt{E-experts} or compressed experts, such that no bubble exists in the schedule, hence $\delta_{v,j}=0, \forall j\in\mathcal{Q}, v\in D_j$.
Combining all the results, we have
\begin{equation}
\label{eq:cmax2}
\begin{aligned}
\texttt{ALG}&\leq\max_{1\leq k\leq M}\left \{ \max \left \{ F_{I}(k), F_{C}(k)\right \} + P(B_{k\rightarrow M}) \right \}\\
&\leq \max_{1\leq k\leq M}\left \{ \max \left \{ I(B_{1\rightarrow k}), \texttt{OPT}+\left (1-\frac{1}{L}\right )\max_{j\in\mathcal{Q}, v\in D_j}\tilde{c}_{v,j}\right \} + P(B_{k\rightarrow M}) \right \}\\
&=\max_{1\leq k\leq M}\left \{ \max \left \{ I(B_{1\rightarrow k}), \texttt{OPT}+\left (1-\frac{1}{L}\right )\max_{j\in\mathcal{Q}, v\in D_j}\left (c+\delta_{v,j}\mathbbm{1}_{X_{b}}\right )\right \} + P(B_{k\rightarrow M}) \right \}\\
&\leq \max_{1\leq k\leq M}\left \{ \max \left \{ I(B_{1\rightarrow k}), \texttt{OPT}+\left (1-\frac{1}{L}\right )\left (c+\rho u\mathbbm{1}_{X_{b}}\right )\right \} + P(B_{k\rightarrow M}) \right \}\\
&\leq \max_{1\leq k\leq M}\left \{ \max \left \{ I(B_{1\rightarrow k}), \texttt{OPT}\right \} +\left (1-\frac{1}{L}\right )\left (c+\rho u\mathbbm{1}_{X_{b}}\right ) + P(B_{k\rightarrow M}) \right \}\\
&= \max_{1\leq k\leq M}\left \{ \max \left \{ I(B_{1\rightarrow k}), \texttt{OPT}\right \} + P(B_{k\rightarrow M}) \right \}+\left (1-\frac{1}{L}\right )\left (c+\rho u\mathbbm{1}_{X_{b}}\right )\\
&\leq \max_{1\leq k\leq M}\left \{ \max \left \{ \texttt{OPT}, \texttt{OPT}\right \} + \texttt{OPT} \right \}+\left (1-\frac{1}{L}\right ) Z\\
&\leq 2\cdot\texttt{OPT} + \left (1-\frac{1}{L}\right )\cdot\texttt{OPT}\\
& = \left (3-\frac{1}{L}\right )\cdot\texttt{OPT},
\end{aligned} 
\end{equation}
where the third inequality holds by Lemma \ref{lem:charge_bound}, the fourth inequality is due to  the fact that $\max\left \{a,b+d\right \}\leq\max\left \{a,b\right \}+d$ holds for every $d\geq 0$, and the fifth inequality follows from Definition \ref{def:critical_path} and Lemma \ref{lem:OPT}.

\end{proof}

\newpage
\section{\textsc{ZipMoE}'s Cache Pool Planning}
\label{sec:code_cache}

This section presents the detailed algorithmic procedures for right-sizing the cache pools for tensors with varying compression states. Let $\mathcal{N}$ denote the set of all expert ranks in a sparse layer ( from 1 to the number of experts per layer ), and $\mathcal{S}$ be any subset of $\mathcal{N}$. 
We denote the \emph{rank-based} expert inclusion probability list as $\textbf{F}_{\mathcal{N}} = [f_1, f_2, \cdots, f_{|\mathcal{N}|}]$, and its normalized counterpart, the \emph{rank-based} expert selection probability list, as $\textbf{P}_\mathcal{N}= [q_1, q_2, \cdots, q_{|\mathcal{N}|}]$.
Procedures for obtaining $\textbf{P}_\mathcal{N}$ from $\textbf{F}_{\mathcal{N}}$ leverages the \emph{modified iterative proportional fitting algorithm}, which is detailed in Appendix \ref{sec:proof_entropy}.
We further define $\textbf{P}_{\mathcal{S}}$ as the partial distribution, represented by a sub-list extracted from $\textbf{P}_{\mathcal{N}}$ according to the indices in $\mathcal{S}$.
We first show how to compute the probability $\Phi_{\mathcal{S}}(h)$ of having exactly $h$ experts sampled from $\mathcal{S}$.

\begin{algorithm}[h]
\caption{Probabilistic Modeling of Cache Hits}
\label{alg:prob_modeling}
\begin{algorithmic}[1]

\State \textbf{Input:} The partial probability distribution $\textbf{P}_{\mathcal{S}}$ for rank-based expert activation.
\State \textbf{Output:} The probability distribution over the number of experts selected from $\mathcal{S}$.

\State Initialize the output distribution $\Phi_{\mathcal{S}}\gets\textbf{0}_{1\times{\left (|\mathcal{S}|+1\right )}}$
\State Initialize boundary condition $\Phi_{\mathcal{S}}(0)\gets1$
\For{$r=1, 2, \cdots, |\mathcal{S}|$}
	\For{$h=|\mathcal{S}|, |\mathcal{S}|-1, \cdots, 1$}
		\State \COMMENT{Update probability using the transition rule.}
		\State $\Phi_{\mathcal{S}}(h)\gets\Phi_{\mathcal{S}}(h)\cdot\left (1-q_r\right ) + \Phi_{\mathcal{S}}(h-1)\cdot q_r$
	\EndFor
	\State $\Phi_{\mathcal{S}}(0)\gets\Phi_{\mathcal{S}}(0)\cdot \left (1-q_r\right )$
\EndFor

\State \textbf{return} Probability distribution $\Phi_{\mathcal{S}}$

\end{algorithmic}
\end{algorithm}

Next, we demonstrate how \textsc{ZipMoE} estimates the sparse layer makespan for any given cache hit pattern $\textbf{h}=\left (h_{\mathcal{F}}, h_{\mathcal{C}}, h_{\mathcal{S}}, h_{\mathcal{E}}\right )$. 
Specifically, the makespan is estimated using average decompression and I/O delays derived from offline profiling.
To circumvent the high computational overhead of fully simulating \emph{Algorithm} \ref{alg:block_priority}, we employ a heuristic approximation: whenever a decompression operation requires an \texttt{E-chunk} read, we penalize its latency by incorporating the corresponding reading delay. 
We then estimate the final makespan as the maximum of two potential bottlenecks: (1) the average computational workload distributed across $L$ threads, and (2) the aggregate I/O workload. 
This approach yields an efficient approximation of the original schedule's performance (excluding token execution).

\begin{algorithm}[h]
\caption{Makespan Estimation}
\label{alg:makespan_estimation}
\begin{algorithmic}[1]

\State \textbf{Input:} Number of activated experts $k$, cache hit pattern $\textbf{h}=\left (h_{\mathcal{F}}, h_{\mathcal{C}}, h_{\mathcal{S}}, h_{\mathcal{E}}\right )$, single \texttt{E-chunk} decompression delay $c$, single \texttt{E-chunk} reading delay $v$, \texttt{SM-chunk} reading delay $u$, number of threads $L$, number of exponent shards in each tensor $K$, number of tensors per expert $n$.
\State \textbf{Output:} Estimation of the sparse layer makespan.

\State \COMMENT{Compute the total I/O workload}
\State $\texttt{n}_{\texttt{SM}}\gets n\cdot\left ( k-\sum_{p\in\left \{ \mathcal{F}, \mathcal{C}, \mathcal{S}\right \}}h_{p} \right )$
\State $\texttt{n}_{\texttt{E}}\gets n\cdot K\cdot\left ( k-\sum_{p\in\left \{ \mathcal{F}, \mathcal{C}, \mathcal{E}\right \}}h_{p} \right )$
\State $T_{I/O}\gets\texttt{n}_{\texttt{SM}}\cdot u+\texttt{n}_{\texttt{E}}\cdot v$

\State \COMMENT{Compute the total decompression workload}
\State $\texttt{n}_{\texttt{D}}\gets n\cdot K\cdot\left (k-h_{\mathcal{F}}\right )$
\State $T_{decomp}\gets \left ( {\texttt{n}_{\texttt{E}}\cdot v + \texttt{n}_{\texttt{D}}\cdot c } \right )/{L}$

\State \textbf{return}  Estimated makespan $\max \left \{ T_{I/O}, T_{decomp} \right \}$

\end{algorithmic}
\end{algorithm}

Finally, the memory for each cache pool can be allocated using our \emph{hierarchical cache planning} outlined by \emph{Algorithm} \ref{alg:pool_planning}. 
Although theoretically feasible, we do not restrict users to partition the memory space for each compression state in practice. 
Instead, we can control the granularity of the cache pools by selecting a subset $\Lambda \in \left \{ \mathcal{F}, \mathcal{C}, \mathcal{S}, \mathcal{E} \right \}$. 
This flexibility allows users to choose the appropriate subset based on the hardware and operating system conditions.

\begin{algorithm}[t]
\caption{Hierarchical Cache Pool Planning}
\label{alg:pool_planning}
\begin{algorithmic}[1]

\State \textbf{Input:} The \emph{rank-based} expert inclusion probability list as $\textbf{F}_{\mathcal{N}}$, candidate set of cache pools  $\Lambda\in\left (h_{\mathcal{F}}, h_{\mathcal{C}}, h_{\mathcal{S}}, h_{\mathcal{E}}\right )$, number of activated experts $k$, single \texttt{E-chunk} decompression delay $c$, single \texttt{E-chunk} reading delay $v$, \texttt{SM-chunk} reading delay $u$, number of threads $L$, number of exponent shards in each tensor $K$, number of tensors per expert $n$, grid size $0<\delta<1$.
\State \textbf{Output:} Memory ratios for each cache pool.

\State Initialization: $\texttt{minCost}\gets \infty$, $\hat{\boldsymbol{\gamma}}:=(\gamma_{\mathcal{F}},\gamma_{\mathcal{C}},\gamma_{\mathcal{S}};\gamma_{\mathcal{E}})\gets(1,0,0,0)$
\State Obtain the expert selection probability list $\textbf{P}_{\mathcal{N}}$ via \emph{modified iterative proportional fitting algorithm} in (\ref{eq:update_weight})

%
			
\For{\textbf{each} feasible memory ratios $\boldsymbol{\gamma}$ such that: $\sum_{p\in\Lambda}{\gamma_p}=1, \gamma_{p'}=0 \text{ for } p'\notin\Lambda $, with step length $\delta$}
			
			\State Compute the maximum number of experts $S_p$ for each cache pool $p\in\Lambda$
			\State \COMMENT{Pack the cache pools into ranks according to the cache hierarchy, compute the distribution via DP}
			\State $u\gets 0$
			\For{$p\in\Lambda\cup\mathcal{M}$}
				\State Compute cache hit distribution $\Phi_p$ on positions $\left \{u+1, u+2, \cdots, u+S_p\right \}$ via \emph{Algorithm} \ref{alg:prob_modeling}
				\State $u\gets u+S_p$
			\EndFor
			\State Compute cache hit distribution $\Phi_\mathcal{N}$ via \emph{Algorithm} \ref{alg:prob_modeling}
			
			\State \COMMENT{Estimate the expected makespan under current allocation}
			\State $\texttt{Cost}\gets0$
			\For{$h_{\mathcal{F}}=0, 1, \cdots, S_\mathcal{F}$}
				\For{$h_{\mathcal{C}}=0, 1, \cdots, S_\mathcal{C}$}
					\For{$h_{\mathcal{S}}=0, 1, \cdots, S_\mathcal{S}$}
						\For{$h_{\mathcal{E}}=0, 1, \cdots, S_\mathcal{E}$}
						
							\State $\texttt{Delay}\gets$Estimate delay using \emph{Algorithm} \ref{alg:makespan_estimation}
							\State $k_{\text{rem}}\gets k-\sum_{p\in\Lambda}h_p$
							\If{$k_{\text{rem}}<0$}
								\State \textbf{break}
							\EndIf
							\State $\texttt{Cost}\gets\texttt{Cost} + \texttt{Delay}\cdot \frac{\Phi_\mathcal{M}(k_{\text{rem}})}{\Phi_\mathcal{N}(k)}\prod_{p\in\Lambda}\Phi_p(h_p)$
							
						\EndFor
					\EndFor
				\EndFor
			\EndFor
			
			\If{\texttt{Cost}$<$\texttt{minCost}}
				\State $\texttt{minCost}\gets\texttt{Cost}$
				\State $\hat{\boldsymbol{\gamma}}\gets\boldsymbol{\gamma}$
			\EndIf
			
\EndFor
			

\State \textbf{return} Optimal memory ratio for each cache pool $\hat{\boldsymbol{\gamma}}$

\end{algorithmic}
\end{algorithm}

\clearpage
\section{Theoretical Analysis of Cache Pool Planning}
\label{sec:proof_entropy}

\begin{theorem}[Restated]
\label{thm:max_ent_restated}
Among all probability distributions over $k$-sized expert subsets that are consistent with the observed individual expert selection counts, the distribution produced by the DP procedure achieves maximum entropy.
\end{theorem}

\noindent
\begin{proof}
To see this, we first construct the mathematical formulation of the entropy maximization problem.
Let $\Omega = 2^{\mathcal{N}}$ be the power set of $\mathcal{N}$ and $\mathcal{P}\left (\mathcal{S}\right )$ be the probability of selecting the set $\mathcal{S}$ of expert IDs.
We have:

\begin{equation}
\label{eq:max_entropy}
\begin{aligned}
\max_{\mathcal{P}} & \sum_{\mathcal{S}\in\Omega}\mathcal{P}\left (\mathcal{S}\right )\log\left (\frac{1}{\mathcal{P}\left (\mathcal{S}\right )}\right )\\
\textrm{s.t.} \quad &C_1:  \mathcal{P}\left (\mathcal{S}\right ) = 0, \forall |\mathcal{S}|\neq k, \\
  &C_2:    \sum_{\mathcal{S}\in\Omega:|\mathcal{S}|=k}\mathcal{P}\left (\mathcal{S}\right )\mathbbm{1}_{i\in\mathcal{S}}={f_i}, \forall i\in \mathcal{N}, \\
  &C_3:  \sum_{\mathcal{S}\in\Omega}\mathcal{P}\left (\mathcal{S}\right )=1. \\
\end{aligned}
\end{equation}

It is straightforward that Eq. (\ref{eq:max_entropy}) is a convex optimization problem, whose Lagrangian can be given by:
\begin{equation}
\label{eq:lagrangian}
\begin{aligned}
\mathcal{L}=-\sum_{\mathcal{S}\in\Omega}\mathcal{P}\left (\mathcal{S}\right )\log\mathcal{P}\left (\mathcal{S}\right ) + \alpha\left (\sum_{\mathcal{S}\in\Omega}\mathcal{P}\left (\mathcal{S}\right )-1\right ) + \sum_{\mathcal{S}\in\Omega:|\mathcal{S}|\neq k}\beta_\mathcal{S}\mathcal{P}\left (\mathcal{S}\right )+\sum_{i\in\mathcal{N}}\lambda_i\left (\sum_{\mathcal{S}\in\Omega:|\mathcal{S}|=k}\mathcal{P}\left (\mathcal{S}\right ) \mathbbm{1}_{i\in\mathcal{S}} - {f_i}\right ),
\end{aligned}
\end{equation}
where $\alpha$, $\beta_\mathcal{S}$, and $\lambda_i$ are the Lagrange multipliers.
The optimal solution $(\boldsymbol{\mathcal{P}}^*, \alpha^*, \boldsymbol{\beta}^*, \boldsymbol{\lambda}^*)$ of Eq. (\ref{eq:max_entropy}) satisfies the Karush-Kuhn-Tucker conditions:
\begin{equation}
\label{eq:KKT}
\begin{aligned}
\left.\frac{\partial\mathcal{L}}{\partial\mathcal{P}\left (\mathcal{S}\right )}\right|_{\boldsymbol{\mathcal{P}}=\boldsymbol{\mathcal{P}}^*}&=-\log\mathcal{P}(\mathcal{S})-1+\alpha+\mathbbm{1}_{|\mathcal{S}|\neq k}+\sum_{i\in\mathcal{S},|\mathcal{S}|=k}\lambda_i=0,\forall\mathcal{S}\in\Omega,\\
\left.\frac{\partial\mathcal{L}}{\partial\alpha}\right|_{\alpha=\alpha^*}&=\sum_{\mathcal{S}\in\Omega}\mathcal{P}\left (\mathcal{S}\right )-1=0,\\
\left.\frac{\partial\mathcal{L}}{\partial\beta_{\mathcal{S}}}\right|_{\boldsymbol{\beta}=\boldsymbol{\beta}^*}&=\mathcal{P}\left (\mathcal{S}\right )=0, \forall\mathcal{S}\in\Omega:|\mathcal{S}|\neq k,\\
\left.\frac{\partial\mathcal{L}}{\partial\lambda_i}\right|_{\boldsymbol{\lambda}=\boldsymbol{\lambda}^*}&=\sum_{\mathcal{S}\in\Omega:|\mathcal{S}|=k}\mathcal{P}\left (\mathcal{S}\right )\mathbbm{1}_{i\in\mathcal{S}}-{f_i}=0, \forall i\in \mathcal{N}, 
\end{aligned}
\end{equation}
solving which yields:
\begin{equation}
\label{eq:maxent_dist}
\begin{aligned}
\mathcal{P}^*\left (\mathcal{S}\right )=\frac{\mathbbm{1}_{|\mathcal{S}|=k}}{\mathcal{Z}}\exp\left ( \sum_{i\in\mathcal{S},|\mathcal{S}|=k}\lambda^*_i \right ), 
\end{aligned}
\end{equation}
where $\mathcal{Z}=\sum_{\mathcal{S}\in\Omega:|\mathcal{S}|=k}\exp\left ( \sum_{i\in\mathcal{S},|\mathcal{S}|=k}\lambda^*_i \right )$ is the normalization factor.

Next, we show that Eq. (\ref{eq:max_entropy}) coincides with the distribution $\mathbb{P}$ resulted from our DP procedure.
Let $X_i$ be the indicator for expert selection, where $X_i=1$ implies that expert $i$ is selected in our sampling procedure,  $X_i=0$ otherwise.
For any set $S$ with cardinality $k$, our \emph{Algorithm}\ref{alg:prob_modeling} samples this set with the following conditional probability:
\begin{equation}
\label{eq:conditional_dist}
\begin{aligned}
\mathbb{P}\left [\mathcal{S}=S \mid \sum_{i\in\mathcal{N}}X_i=k\right ]=\frac{\mathbb{P}\left [\mathcal{S}=S, \sum_{i\in\mathcal{N}}X_i=k\right ]}{\mathbb{P}\left [\sum_{i\in\mathcal{N}}X_i=k\right ]}.
\end{aligned}
\end{equation}
Let $q_i$ be the selection probability for expert $i$.
Since $|S|=k$, we have:
\begin{equation}
\label{eq:conditional_up}
\begin{aligned}
\mathbb{P}\left [\mathcal{S}=S, \sum_{i\in\mathcal{N}}X_i=k \right ] &= \left (\prod_{i\in S}q_i\right )\left (\prod_{i\in \mathcal{N}\setminus S}\left (1-q_i\right )\right )\\
&=\frac{\left (\prod_{i\in S}q_i\right )\left (\prod_{i\in \mathcal{N}\setminus S}\left (1-q_i\right )\right )\left (\prod_{i\in S}\left (1-q_i\right )\right )}{\prod_{i\in S}\left (1-q_i\right )}\\
&=\left (\prod_{i\in S}\frac{q_i}{1-q_i}\right )\left (\prod_{i\in \mathcal{N}}\left (1-q_i\right )\right )\\
&=\Gamma\left (\prod_{i\in S}\frac{q_i}{1-q_i}\right ),
\end{aligned}
\end{equation}
where $\Gamma=\prod_{i\in \mathcal{N}}\left (1-q_i\right )$ is constant.
On the other hand,
\begin{equation}
\label{eq:conditional_down}
\begin{aligned}
\mathbb{P}\left [\sum_{i\in\mathcal{N}}X_i=k \right ] &= \sum_{J\in\Omega:|J|=k}\mathbb{P}\left [[\mathcal{S}=J, \sum_{i\in\mathcal{N}}X_i=k \right ]=\Gamma\sum_{J\in\Omega:|J|=k}\left (\prod_{i\in J}\frac{q_i}{1-q_i}\right ),
\end{aligned}
\end{equation}
where the last equality follows from Eq. (\ref{eq:conditional_up}).
Combining Eq. (\ref{eq:conditional_up}) and Eq. (\ref{eq:conditional_down}), we have:
\begin{equation}
\label{eq:conditional_dist}
\begin{aligned}
\mathbb{P}\left [\mathcal{S}=S \mid \sum_{i\in\mathcal{N}}X_i=k\right ]&=\frac{\mathbb{P}\left [\mathcal{S}=S, \sum_{i\in\mathcal{N}}X_i=k\right ]}{\mathbb{P}\left [\sum_{i\in\mathcal{N}}X_i=k\right ]}=\frac{\prod_{i\in S}\frac{q_i}{1-q_i}}{\sum_{J\in\Omega:|J|=k}\left (\prod_{i\in J}\frac{q_i}{1-q_i}\right )}\\
&=\frac{\mathbbm{1}_{|S|=k}\exp\left (\sum_{i\in S}\lambda_i\right )}{\sum_{J\in\Omega:|J|=k}\exp\left (\sum_{j\in J}\lambda_j\right )},
\end{aligned}
\end{equation}
where the last equality follows by the definition $\lambda_i:=\log\left (\frac{q_i}{1-q_i}\right )$.
Thus, $\mathbb{P}$ and $\mathcal{P}$ belong to the same parametric family of distributions.
To show that $\mathbb{P}$ and $\mathcal{P}$ are essentially identical, we must find a set of sampling probabilities $q^*_i$ for each expert rank $i$ such that constraint $C_2$ in Eq. (\ref{eq:max_entropy}) is satisfied.
This problem belongs to the class of weighted finite population sampling problems. 
It was shown by \cite{BM94} that for any feasible set of inclusion probabilities $\{f_i\}_{i=1}^N$, there exists a unique set of positive weights $\{w^*_i\}_{i=1}^N$ (up to a scaling factor) such that the resulting distribution satisfies:(i) ${P}(\mathcal{S}) \propto \prod_{i\in\mathcal{S}}w^*_i$;(ii) $\sum_{\mathcal{S} \ni i, |\mathcal{S}|=k}{P}(\mathcal{S}) = f_i, \forall i\in\mathcal{N}$;and (iii) $\sum_{i\in\mathcal{N}}f_i=k$.
Such set of $w^*_i$ can be found via a \emph{modified iterative proportional fitting algorithm} \cite{BM94}, which we restate as follows.
We first define the \emph{elementary symmetric polynomial} with respect to integer $n$ and set $\mathcal{C}$ as:
\begin{equation}
\label{eq:symmetric_poly}
\begin{aligned}
\mathcal{R}\left (n,\mathcal{C}\right ):=\sum_{S\subseteq\mathcal{C}, |S|=n}\prod_{j\in S}w_j.
\end{aligned}
\end{equation}
Let the iteration be indexed by $t$. 
Starting with $w^{(0)}_i=f_i$, we update the weights according to: $w^{(t+1)}_i \gets w^{(t)}_i \cdot \frac{f_i}{f^{(t)}_i}$, where
\begin{equation}
\label{eq:update_weight}
\begin{aligned}
f^{(t)}_i:=\left.w^{(t)}_i\frac{\mathcal{R}\left (k-1,\mathcal{N}\setminus\left \{i\right \} \right)}{\mathcal{R}\left (k,\mathcal{N}\right )}\right|_{\boldsymbol{w}=\boldsymbol{w}^{(t)}},\forall i\in \left \{1,2,\cdots, |\mathcal{N}|\right \}.
\end{aligned}
\end{equation}
Such an iteration converges monotonically and geometrically to $w^*_i$.
Finally, we can set $w^*_i = \exp(\lambda^*_i) = \frac{q^*_i}{1-q^*_i}$, or equivalently $q^*_i = \frac{w^*_i}{1+w^*_i}$. 
The resulting set of $\{q^*_i\}$ satisfies constraint $C_2$ of Eq. (\ref{eq:max_entropy}) and, therefore, produces a distribution that coincides with the maximum entropy distribution $\mathcal{P}^*(\mathcal{S})$.

\end{proof}

\newpage
\section{Extended Results}
\label{sec:extended_results}

\subsection{Offline Initialization}

We report the offline initialization time in Tab. \ref{tab:offline_compression}.
Overall, offline initialization spans from 6 minutes to 20 minutes, depending on specific models and compressor configurations. 
As this process is performed only once per model, the cost is practically affordable. 
Regarding memory overhead, ZipMoE employs parallel and asynchronous compression in its offloading engine. 

%

\begin{table}[h]
\centering
\caption{Offline compression latency for different models and compressors.}
\label{tab:offline_compression}
\begin{tabular}{lSS} %
\toprule
\textbf{Model} & \multicolumn{2}{c}{\textbf{Compression Time (s)}} \\
\cmidrule(lr){2-3}
& {\textbf{LZ4HC}} & {\textbf{ZSTD}} \\ 
\midrule
DeepSeekV2-Lite 16B             & 760.02  & 427.99 \\
Qwen1.5-MoE 14B                 & 680.89  & 384.05 \\
SwitchTransformers-Large-128 26B & 1186.96 & 594.68 \\
\bottomrule
\addlinespace[1ex]
\multicolumn{3}{l}{}
\end{tabular}
\end{table}

In addition to the memory required for model shards, extra RAM is required to buffer batches of tensors and support the compression pipeline. In practice, we observe less than 1 GB of additional memory overhead while achieving high compression throughput.

\subsection{Sensitivity Analysis}

We analyze the sensitivity of the cache-planning module under distribution shifts in expert activation probabilities. 
Specifically, we inject controlled distribution shifts into the cache-planning procedure by introducing a shift factor $\alpha$ to smooth the historical rank-based expert activation count list $\textbf{F}_\mathcal{N}$. 
The shifted activation count list $\tilde{\textbf{F}}_\mathcal{N}$ is defined as
\[
\tilde{\textbf{F}}_\mathcal{N}=(1-\alpha)\times\textbf{F}_\mathcal{N}+\alpha\times\mathbb{E}(\textbf{F}_\mathcal{N}).
\]
We then use the shifted list $\tilde{\textbf{F}}_\mathcal{N}$ as the input to the cache-planning module, thereby introducing a discrepancy between the planning distribution and the actual workload distribution. 
The experimental results are presented in Tab.~\ref{tab:sensitivity_workload}.

%

\begin{table}[h]
\centering
\caption{Sensitivity against workload distribution shift.}
\label{tab:sensitivity_workload}
\setlength{\tabcolsep}{14pt} %
\begin{tabular}{ccccc}
\toprule
 & \multicolumn{2}{c}{\textbf{TTFT}} & \multicolumn{2}{c}{\textbf{TPOT}} \\
\cmidrule(lr){2-3} \cmidrule(lr){4-5}
Shift Factor \textbf{$\alpha$} & \textbf{Latency (s)} & \textbf{Slowdown} & \textbf{Latency (s)} & \textbf{Slowdown} \\ 
\midrule
0.0 & 2.1783 & 1.000$\times$ & 0.4576 & 1.000$\times$ \\
0.2 & 2.3055 & 1.058$\times$ & 0.4709 & 1.029$\times$ \\
0.4 & 2.3634 & 1.085$\times$ & 0.4762 & 1.041$\times$ \\
0.6 & 2.3616 & 1.084$\times$ & 0.4777 & 1.044$\times$ \\
0.8 & 2.5285 & 1.161$\times$ & 0.4989 & 1.090$\times$ \\
1.0 & 2.5442 & 1.168$\times$ & 0.5011 & 1.095$\times$ \\
\bottomrule
\end{tabular}
\end{table}

The results demonstrate that \textsc{ZipMoE}'s cache-planning module remains highly robust under workload distribution shifts. 
Even under the largest shift setting, the system incurs only a 16.8\% increase in TTFT and a 9.5\% increase in TPOT.

\begin{figure*}[t]
    \centering
    \includegraphics[width=0.88\textwidth]{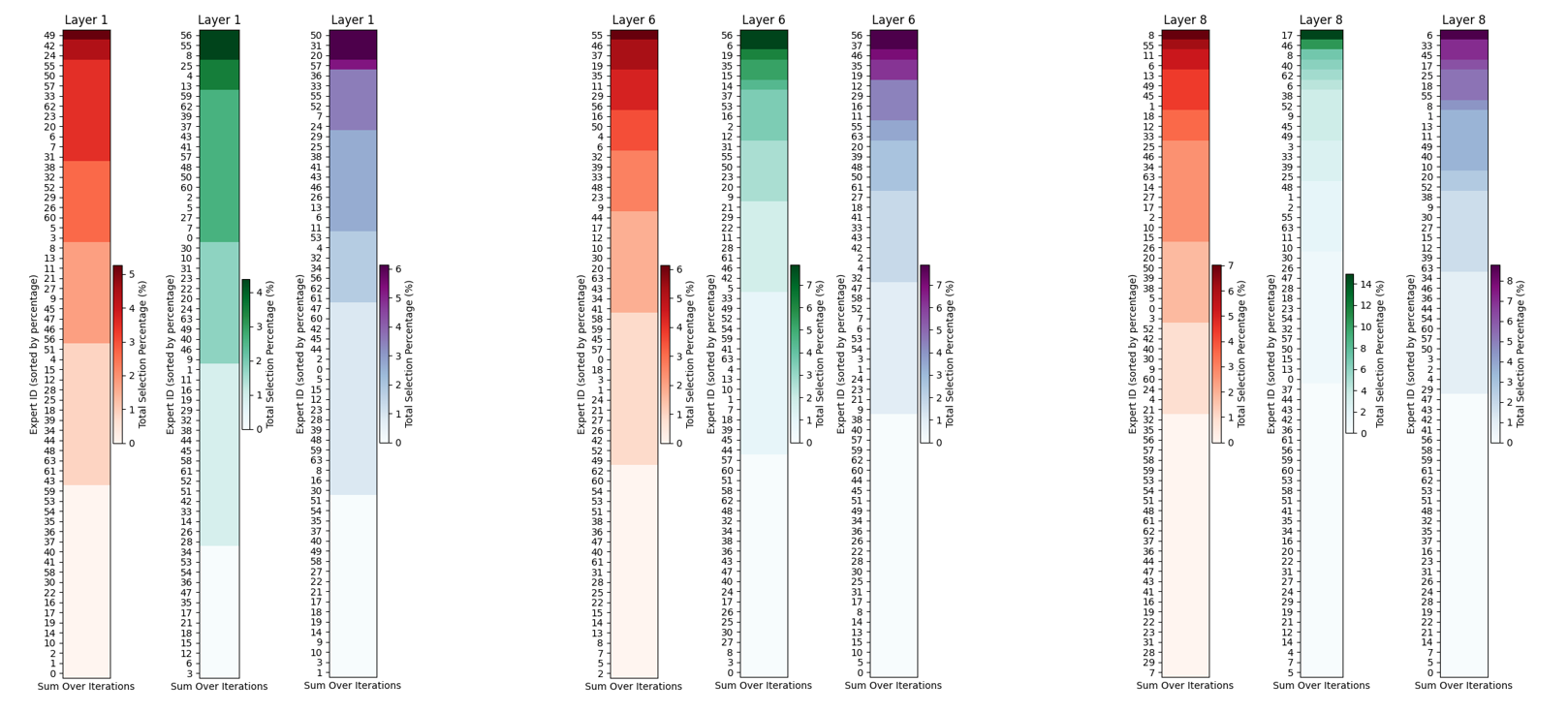}
    \caption{Rank-based distribution of expert activations. Different colors correspond to different prompts.}
    \label{fig:trace}
\end{figure*}

Here, we explain the rationale behind the robustness of this module. In practical deployments, while the specific IDs of activated experts may change across different prompts, the skewness (shape) of the rank-based activation probability distribution remains highly stable. 
As the cache-planning of \textsc{ZipMoE} relies on rank-based popularity rather than absolute expert IDs, it successfully captures the invariant skewness. Hence, the planning is robust.
We validated this by collecting inference traces using DeepSeekV2-Lite 16B model on the ShareGPT corpus.
The trace distribution is illustrated in Fig. \ref{fig:trace}.

\newpage

\newcolumntype{Y}{>{\centering\arraybackslash}X}

\begin{table}[h]
\centering
\caption{TTFT performance (in seconds) of Qwen1.5-MoE 14B under different memory footprints.}
\label{tab:ttft_comparison}
\begin{tabularx}{\linewidth}{l YYYYYY}
\toprule
\textbf{Baseline \textbackslash RAM} & \textbf{5GB} & \textbf{10GB} & \textbf{15GB} & \textbf{20GB} & \textbf{25GB} & \textbf{30GB} \\ 
\midrule
ZipMoE  & 3.1312   & 2.4597  & 2.0486  & 1.6392  & 1.3020  & 1.2303  \\
FineMoE  & 20.6469 & 20.6426 & 20.7792 & 23.6066 & 23.4409 & 19.4810 \\
\bottomrule
\end{tabularx}
\end{table}

\begin{table}[h]
\centering
\caption{TPOT performance (in seconds) of Qwen1.5-MoE 14B under different  memory footprints.}
\label{tab:tpot_comparison}
\begin{tabularx}{\linewidth}{l YYYYYY}
\toprule
\textbf{Baseline \textbackslash RAM} & \textbf{5GB} & \textbf{10GB} & \textbf{15GB} & \textbf{20GB} & \textbf{25GB} & \textbf{30GB} \\ 
\midrule
ZipMoE  & 0.4082  & 0.3688   & 0.3331   & 0.3060   & 0.2644   & 0.2395  \\
FineMoE  & 20.4272 & 20.4053  & 20.0166 & 23.1935  & 22.9447  & 19.4125 \\
\bottomrule
\end{tabularx}
\end{table}

\subsection{More Baseline Comparison}

We compare \textsc{ZipMoE} with a more recent baseline \emph{FineMoE}~\cite{FineMoE} in Tab. \ref{tab:ttft_comparison} and Tab. \ref{tab:tpot_comparison}.
Somewhat surprisingly, although \emph{FineMoE} achieves strong performance in cloud environments, its efficiency degrades significantly on edge devices. This is due to miss prefetching quickly saturates PCIe bandwidth of edge devices, which is more likely to occur in single-batch on-device workloads.

\newpage
\subsection{Comparison between Different Compressors}

\begin{table}[t]
\centering
\caption{Comparison of different compression techniques across varying memory footprints.}
\label{tab:compression_comparison}
\begin{tabular}{lcccccc}
\toprule
\textbf{Memory} & \multicolumn{3}{c}{\textbf{TTFT (s)}} & \multicolumn{3}{c}{\textbf{TPOT (s)}} \\
\cmidrule(lr){2-4} \cmidrule(lr){5-7}
\textbf{Footprint (GB)} & LZ4HC & ZSTD & $\Delta$ & LZ4HC & ZSTD & $\Delta$ \\
\midrule
10 & 3.0456 & 3.3183 & +0.2727 & 0.5484 & 0.5980 & +0.0497 \\
15 & 2.6663 & 2.8940 & +0.2277 & 0.5088 & 0.5393 & +0.0305 \\
20 & 2.1783 & 2.4235 & +0.2452 & 0.4576 & 0.4889 & +0.0313 \\
25 & 1.7508 & 1.9697 & +0.2188 & 0.4115 & 0.4219 & +0.0104 \\
30 & 1.5525 & 1.6544 & +0.1018 & 0.3666 & 0.3614 & $-0.0052$ \\
\bottomrule
\end{tabular}
\end{table}

We compare the system performance using different compressor backends in Tab. \ref{tab:compression_comparison}.
Overall, the choice between LZ4HC and ZSTD as the compression backend does not lead to significant performance differences. LZ4HC shows a slight advantage due to its better balance between compression ratio and decompression speed.

That said, \textsc{ZipMoE}’s current implementation exposes an open API that supports a wide range of compressor backends. This flexibility allows users to select the most suitable compressor for their target edge devices.


\end{document}